\documentclass[conference]{IEEEtran}
\IEEEoverridecommandlockouts
\usepackage{cite}
\usepackage{amsmath,amssymb,amsfonts}
\usepackage{multirow}
\usepackage{booktabs}       
\newtheorem{theorem}{Theorem}
\usepackage[caption=false]{subfig}
\usepackage{algorithmic}
\usepackage{graphicx}
\usepackage{hyperref}       
\usepackage{textcomp}
\usepackage{xcolor}
\def\BibTeX{{\rm B\kern-.05em{\sc i\kern-.025em b}\kern-.08em
    T\kern-.1667em\lower.7ex\hbox{E}\kern-.125emX}}
\begin{document}

\title{Low-complexity Deep Video Compression with A Distributed Coding Architecture}

\author{\IEEEauthorblockN{Xinjie Zhang, Jiawei Shao, and Jun Zhang}
\IEEEauthorblockA{\textit{Department of Electronic and Computer Engineering} \\
\textit{The Hong Kong University of Science and Technology}\\
Hong Kong, China \\
\{xinjie.zhang, jiawei.shao\}@connect.ust.hk, eejzhang@ust.hk}
}

\maketitle

\begin{abstract}
Prevalent predictive coding-based video compression methods rely on a heavy encoder to reduce temporal redundancy, which makes it challenging to deploy them on resource-constrained devices. Since the 1970s, distributed source coding theory has indicated that independent encoding and joint decoding with side information (SI) can achieve high-efficient compression of correlated sources. This has inspired a \textit{distributed coding} architecture aiming at reducing the encoding complexity. However, traditional distributed coding methods suffer from a substantial performance gap to predictive coding ones. Inspired by the great success of learning-based compression, we propose the first end-to-end distributed deep video compression framework to improve the rate-distortion performance. A key ingredient is an effective SI generation module at the decoder, which helps to effectively exploit inter-frame correlations without computation-intensive encoder-side motion estimation and compensation. Experiments show that our method significantly outperforms conventional distributed video coding and H.264. Meanwhile, it enjoys $6\sim7\times$ encoding speedup against DVC \cite{lu2019dvc} with comparable compression performance. Code is released at {\color{magenta} \href{https://github.com/Xinjie-Q/Distributed-DVC}{https://github.com/Xinjie-Q/Distributed-DVC}}.
\end{abstract}


\section{Introduction}
\label{sec:intro}  
\begin{figure*}[t]
  \centering
  \subfloat[The predictive coding architecture used by standard and most learning-based video codecs.] 
  {\label{subfig:AVC_HEVC}
  \includegraphics[scale=0.3]{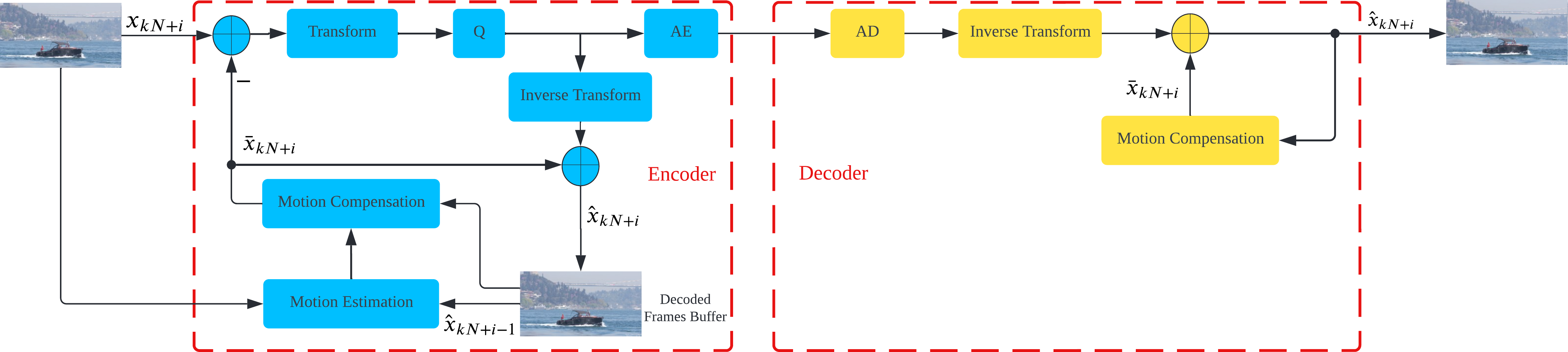}
  }\\
  \subfloat[The proposed distributed deep video coding architecture.]
  {\label{subfig:distributed_dvc}
  \includegraphics[scale=0.3]{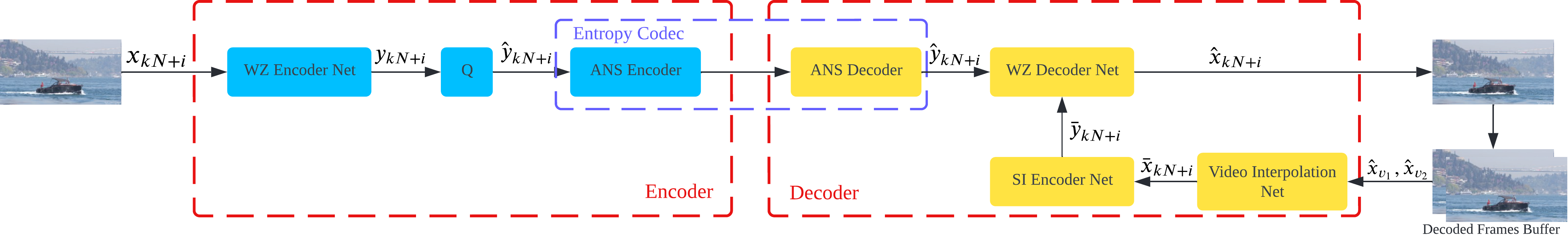}
  }
  \caption{Comparison between predictive coding methods and the proposed distributed deep video coding framework.}
  \label{fig:AVC_DistriVC_architectures}
  \vspace{-0.4cm}
\end{figure*}

Given the ubiquity and popularity of various video applications, deep learning (DL)-based video compression \cite{lu2019dvc, lu2020end, agustsson2020scale, hu2021fvc, li2021deep, li2022hybrid} has attracted increasing attention due to their superior performance over traditional methods \cite{wiegand2003overview, sullivan2012overview}. Most methods adopt a predictive coding architecture inherited from standard video codecs. As illustrated in \figurename~\ref{fig:AVC_DistriVC_architectures}(a), this architecture applies a computation-intensive motion compensation prediction loop at the encoder to explicitly reduce temporal redundancy between frames. According to \tableautorefname~\ref{table:complexity_analysis}, the motion-related operations (\textit{i.e.}, estimation, compensation and compression) account for about 90\% and 65\% computation complexity in DVC \cite{lu2019dvc} and DCVC \cite{li2021deep}, two representative examples of DL-based video codecs. Therefore, these DL-based methods are typically characterized by a heavy encoder and a relatively simple decoder, making them suitable for broadcast-oriented applications (\textit{e.g.}, video streaming \cite{fan2019survey} and Blu-Ray discs \cite{miyagawa2014overview}), where the videos are encoded once and decoded many times.

\begin{table*}
\small
  \caption{Complexity of key components for learning-based video encoders on 1080p videos running on an Intel Xeon Gold 6230R processor with a base frequency 2.10GHz and a single CPU core.}
  \label{table:complexity_analysis}
  \vspace{-0.2cm}
  \centering
  \begin{tabular}{c|ccc|c|c}
    \hline
    \multirow{2}*{Latency/FLOPs} & \multicolumn{3}{c|}{Motion} & Residual/WZ & \multirow{2}*{Total}  \\
    & Estimation & Compensation & Compression & Compression\\
    \hline
    \multirow{2}*{DCVC \cite{li2021deep}}  & 13.02s & 12.74s &24.15s & 29.28s & 79.18s\\
     & 1253.65G & 732.83G & 1049.95G & 1557.32G & 4593.79G\\
    \hline
    \multirow{2}*{DVC \cite{lu2019dvc}}  & 13.53s & 15.90s & 12.60s & 3.13s & 45.16s \\
     & 1253.65G & 783.14G & 635.04G & 187.3G & 2859.12 G \\
    \hline
    \multirow{2}*{Proposed}  & 0 &0 & 0 & 6.68s & 6.68s \\
     & 0 & 0 & 0 & 500.45G & 500.45G \\    
    \hline
  \end{tabular}
  \vspace{-0.4cm}
\end{table*}

Recently there is an upsurge in uplink-based video applications such as video surveillance \cite{elharrouss2021review} and multi-view image acquisition \cite{zhang2023ldmic}, where the video encoder is deployed on a device with limited computational resource and power supply. Such scenarios demand low-complexity and low-power video encoders, rendering existing DL-based video codecs unsuitable. This calls for a radical change in the video coding architecture. Inspired by Slepian-Wolf (SW) \cite{Slepian1973} and Wyner-Ziv (WZ) \cite{Wyner1976} theorems\footnote{More details about these two theorems are provided in Appendix \ref{appendix:sw_and_wz_theorem}.} developed in the 1970s on distributed source coding, \textit{distributed video coding}\footnote{Distributed video coding has been called as DVC in literatures, but as DVC is more often used to refer to deep video compression recently, we call distributed video coding as WZ video coding in this paper. More details about WZ video coding are provided in Appendix \ref{appendix:classic_wz_video_coding}.}, also known as WZ video coding \cite{Girod2005}, has emerged as a promising solution to complement existing video compression methods for multimedia applications. 

Nevertheless, it is non-trivial to build a high-efficient and low-complexity practical video compression system by directly applying WZ video coding. \textbf{Firstly}, existing WZ video codecs only achieve the rate-distortion (RD) performance similar with H.264-intra coding \cite{kodavalla2011chroma, kodavalla2012chroma}. There is a large gap to popular standard video codecs, which is predominantly caused by the poor quality of side information (SI) at the decoder \cite{dufaux2010distributed}. It is unclear how to produce high-quality SI and effectively exploit SI at the decoder to approach the performance of predictive coding. \textbf{Secondly}, there is a feedback channel from the decoder to encoder in classical SW codecs \cite{aaron2002wyner, dash2018decoder, dash2019decoder}. The decoder has to request additional parity bits repeatedly until the decoding procedure is successful, resulting in a large decoding delay. Furthermore, it demands an extra frame buffer to store the encoded stream, which consumes high memory at the encoder, making it unsuitable for mobile devices and cameras. To guarantee user experience for video applications, it is critically important to develop resource-friendly video compression methods. 

In this paper, we design the first end-to-end distributed deep video coding (Distributed DVC) system, which boosts the coding gains to match the ones with the predictive coding architecture by exploiting the advantages of deep neural networks (DNNs) in nonlinear transform and end-to-end optimization, while alleviating high encoding overheads of learning-based video codecs by removing the computation-intensive motion operations from the encoder side. The main contributions of this paper are summarized below. \textbf{Firstly}, our Distributed DVC enables low-complexity encoders, which are desirable for uplink-based video applications with resource-constrained devices. While distributed source coding theory has inspired such an architecture decades ago, there is a lack of practical methods that can achieve RD performance close to the predictive coding architecture, and our study fills the gap. \textbf{Secondly}, an \textit{SI generation module} at the decoder is proposed to explicitly capture the temporal correlations between frames for reconstructing more informative frames and an encoder-decoder joint training strategy is introduced to implicitly make latent representation more compact. Furthermore, we use a \textit{channel-wise auto-regressive entropy model} \cite{minnen2020channel} to provide accurate probability distribution for the latent representation during entropy coding, which effectively reduces the statistical redundancy. Meanwhile, it replaces the SW codec and avoids the problems brought by the feedback channel in traditional WZ video coding. \textbf{Finally}, extensive experimental results demonstrate that our framework provides up to 10 dB gains over the traditional WZ video codec \cite{kodavalla2012chroma} in PSNR and outperforms H.264. Compared with H.265, our method achieves higher coding gains in most of bit ranges when measured by MS-SSIM. Meanwhile, it reduces about 85\% and 90\% encoding latency against DVC \cite{lu2019dvc} and DCVC \cite{li2021deep}, respectively.

\section{Related Works}
\label{related_works}
\textbf{Learning-based Compression.} 
Compared with standard codecs, recent learning-based image and video compression methods have made significant progresses in terms of RD performance. Most deep image compression methods employ an auto-encoder style network with various types of entropy models for high compression performance \cite{balle2017end, balle2018variational, minnen2018joint, minnen2020channel, cheng2020learned, he2021checkerboard}. In addition, other techniques such as generalized divisive normalization (GDN) \cite{balle2016density}, latent residual prediction and round-based training \cite{minnen2020channel} have been proposed to improve the coding gains. We consider these existing works to be important building blocks for our framework.

Existing learning-based video compression methods \cite{lu2019dvc, agustsson2020scale, hu2021fvc, li2021deep} mainly follow the predictive coding architecture. They suffer from high encoding complexity due to heavy motion-related operations. By contrast, our method achieves a better trade-off between RD performance and encoding complexity, which is suitable for computation resource-constrained situations and complements current learning-based video compression approaches. Moreover, different from orthogonal works that apply model quantization to reduce the complexity of learning-based image codecs \cite{hong2020efficient, sun2021end, jia2022fpx}, our work aims at reducing the encoding complexity by changing the coding architecture.

\textbf{Distributed Video Coding.} 
Over the past few decades, various handcrafted tools have been proposed to improve the compression performance of WZ video coding, including the application of different transformation \cite{aaron2004transform}, more accurate correlation noise estimation \cite{brites2008correlation} and the refinement of SI \cite{artigas2007discover, ren2011new}. Recent works have introduced DNNs to boost the performance \cite{bhagath2016low, dash2018decoder, dash2019decoder}. However, only some of the main modules are replaced by DNNs, which hinders WZ video codecs from enjoying the benefits of end-to-end optimization. By contrast, all key components in our proposed framework are implemented with DNNs and jointly optimized to significantly improve the RD performance. 

\section{Proposed Method}
\label{proposed_method}
\textbf{Notations.} 
Let $\mathcal{X}=\{x_1,x_2,...\}$ denote the original video sequence. WZ video coding applies an adaptive or fixed frame separator to split the video sequence into key frames and WZ frames. For simplicity, we assume a fixed size of group of pictures (GOP) as $N$. In this case, $x_{kN+1}$ represents a key frame of the video sequence, and the other frames $x_{kN+i}$ are WZ frames, where $k=\{0,1,2,...\}$ and $i=\{2,3,...,N\}$. $\hat{x}_{kN+i}$ denotes the reconstructed WZ frame. In order to exploit the temporal correlation, an SI frame $\bar{x}_{kN+i}$ is generated using two previous decoded frames $\hat{x}_{v_1}$ and $\hat{x}_{v_2}$ at the decoder, where $v_1$ and $v_2$ denote the index of the reference frames. Transform coding can be used to improve the compression efficiency. In such case, the original frame $x_{kN+i}$ and SI frame $\bar{x}_{kN+i}$ are transformed to $y_{kN+i}$ and $\bar{y}_{kN+i}$, respectively. $\hat{y}_{kN+i}$ is the quantized version of $y_{kN+i}$.  

\begin{figure}[t]
\centering
\includegraphics[scale=0.5]{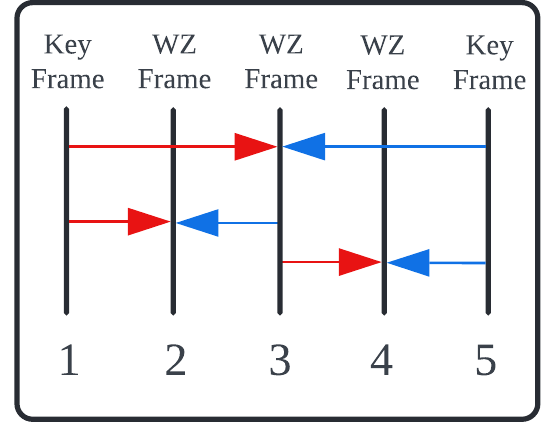}
  \vspace{-0.3cm}
\caption{The hierarchical frame interpolation order.}
\label{fig:interpolation_order}
\vspace{-0.4cm}
\end{figure}

\begin{figure*}[t]
  \centering
  \includegraphics[scale=0.225]{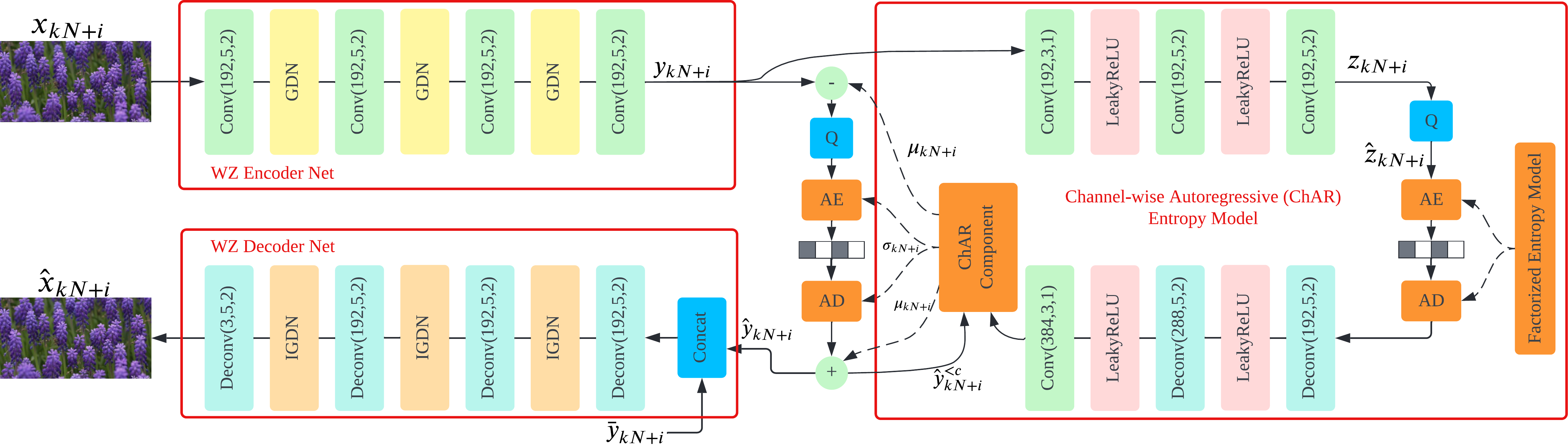}
  \vspace{-0.3cm}
  \caption{Our proposed WZ encoder-decoder network. Convolution parameters are formatted as (the number of filters, kernel size, stride). AE and AD denote ANS encoder and decoder, respectively.}
  \label{fig:WZ_auto_encoder}
  \vspace{-0.4cm}
\end{figure*}

\subsection{Overview of the Proposed Method}
\figurename~\ref{fig:AVC_DistriVC_architectures}(b) presents a high-level overview of our framework. The following describes details about each step and highlights key differences from the classic WZ video codec \cite{kodavalla2012chroma}:

\textbf{Step 1. Transformation and quantization.} 
We replace the linear transformation in \cite{kodavalla2012chroma} with a non-linear WZ encoder network. The input WZ frame $x_{kN+i}$ is mapped to the representation $y_{kN+i}$. Then $y_{kN+i}$ is quantized to $\hat{y}_{kN+i}$. To enable end-to-end training, we adopt the quantization method in \cite{minnen2020channel}. Details are presented in Sec. \ref{subsec:encoder_decoder}.

\textbf{Step 2. Entropy encoding.} 
Instead of using an SW codec to compress the quantized WZ representation, we adopt an asymmetrical numeral systems (ANS) encoder \cite{duda2013asymmetric} to encode the quantized WZ representation $\hat{y}_{kN+i}$ into a bitstream at the inference stage, where the probability distribution of $\hat{y}_{kN+i}$ is estimated by a channel-wise auto-regressive (ChAR) entropy model. Further details are given in Sec. \ref{subsec:encoder_decoder}.

\textbf{Step 3. Side information generation.} 
An optical flow video interpolation network in \cite{huang2020rife} is adopted to generate the current SI frame $\bar{x}_{kN+i}$ based on two previous decoded frames $\hat{x}_{v_1}$ and $\hat{x}_{v_2}$ with the hierarchical interpolation order shown in \figurename~\ref{fig:interpolation_order}. Furthermore, an SI encoder is designed to transform $\bar{x}_{kN+i}$ to the SI representation $\bar{y}_{kN+i}$. More information is provided in Sec. \ref{subsec:side_information}. 

\textbf{Step 4. Entropy decoding.} 
The decoder receives the bitstream from the encoder and performs ANS decoding to reconstruct the quantized WZ representation $\hat{y}_{kN+i}$. Compared with SW decoding, the entropy decoding is more efficient and removes the feedback channel of traditional WZ video coding.

\textbf{Step 5. Inverse transformation.} 
We concatenate the quantized WZ representation $\hat{y}_{kN+i}$ and SI representation $\bar{y}_{kN+i}$, and feed them into the decoder network to reconstruct the WZ frame $\hat{x}_{kN+i}$, which is different from using a predefined quantization table to perform the merge operation and reconstruction process in \cite{kodavalla2012chroma}. Details are presented in Sec. \ref{subsec:encoder_decoder}.

\subsection{WZ Encoder and Decoder Networks}
\label{subsec:encoder_decoder}
To facilitate the compression of the WZ frame $x_{kN+i}$, we take inspiration from deep image compression \cite{balle2018variational,minnen2018joint} and design a CNN-based WZ encoder-decoder network. As illustrated in \figurename~\ref{fig:WZ_auto_encoder}, given an input frame $x_{kN+i}$, the WZ encoder generates the representation $y_{kN+i}$ that is then quantized to $\hat{y}_{kN+i}$. The WZ decoder receives the quantized representation from the encoder and reconstructs the WZ information $\hat{x}_{kN+i}$ with the aid of the SI representation $\bar{y}_{kN+i}$ from the SI generation module. Moreover, we employ a ChAR entropy model \cite{minnen2020channel} to estimate the probability distribution of $y_{kN+i}$, where a hyper prior entropy model \cite{balle2018variational} generates the hyper representation $\hat{z}_{kN+i}$ to capture the spatial redundancies among the elements of $\hat{y}_{kN+i}$, and a ChAR component exploits the correlations among the channels of $\hat{y}_{kN+i}$. These two ingredients produce the mean and scale parameters of a conditional Gaussian entropy model. In addition, since the quantization operation is not differentiable, we apply the mixed quantization method in \cite{minnen2020channel} to enable optimization via stochastic gradient descent. Specifically, $y_{kN+i}$ with an additive uniform noise is fed into the entropy model for bitrate estimation, while the rounded representation $\hat{y}_{kN+i}$ based on straight-through estimation is used as the input to the decoder for reconstruction.


\subsection{Side Information Generation Network}
\label{subsec:side_information}
\begin{figure}[t]
  \centering
  \includegraphics[scale=0.35]{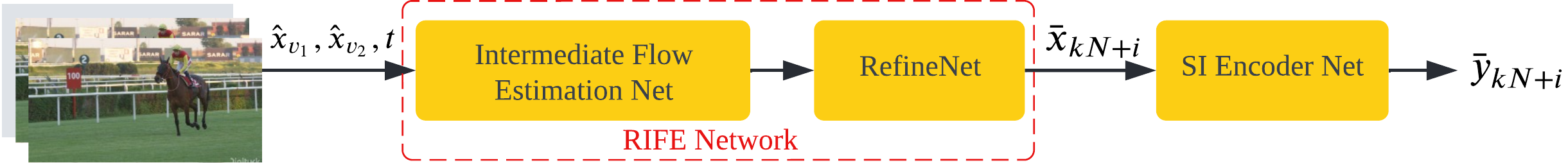}
  \caption{Side information generation network.}
  \label{fig:side_information}
  \vspace{-0.5cm}
  \end{figure}

As shown in \figurename~\ref{fig:side_information}, our SI generation network is composed of two parts: an optical flow-based video interpolation network and an SI encoder. We adopt the RIFE network \cite{huang2020rife} for video interpolation that allows arbitrary time-step frame interpolation with two input frames. Specifically, given two previous decoded frames $\hat{x}_{v_1}, \hat{x}_{v_2}$ and the corresponding time step $t$ $(0\leq t \leq 1)$, the intermediate flow estimation network estimates the motion information and produces a coarse interpolated frame. Then the RefineNet reduces artifacts and refines the high-frequency area to create the current predicted frame $\bar{x}_{kN+i}$, which is expected to be as close to the current frame $x_{kN+i}$ as possible. Next, we utilize the SI encoder with the same network as the WZ encoder described in Sec. \ref{subsec:encoder_decoder} to produce the SI representation $\bar{y}_{kN+i}$, which is then concatenated with the WZ representation $\hat{y}_{kN+i}$ to exploit temporal correlations for generating higher quality frame. More details of the network are given in Appendix~\ref{appendix:network_arch}. 


\begin{figure*}[t]
  \centering
  \subfloat 
  {\label{subfig:uvg_psnr}
  \includegraphics[scale=0.215]{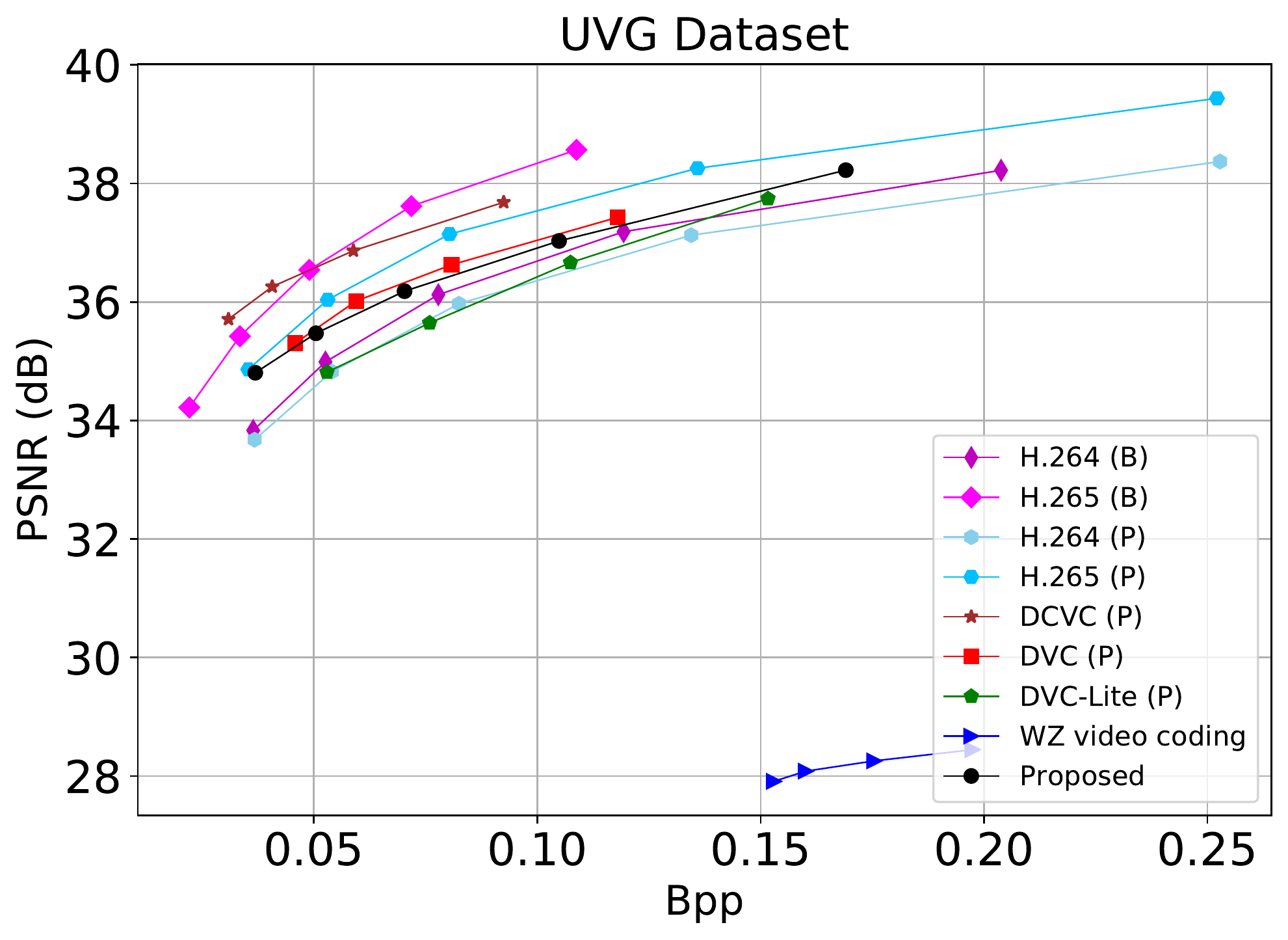}
  }
  \subfloat
  {\label{subfig:uvg_ms_ssim}
  \includegraphics[scale=0.215]{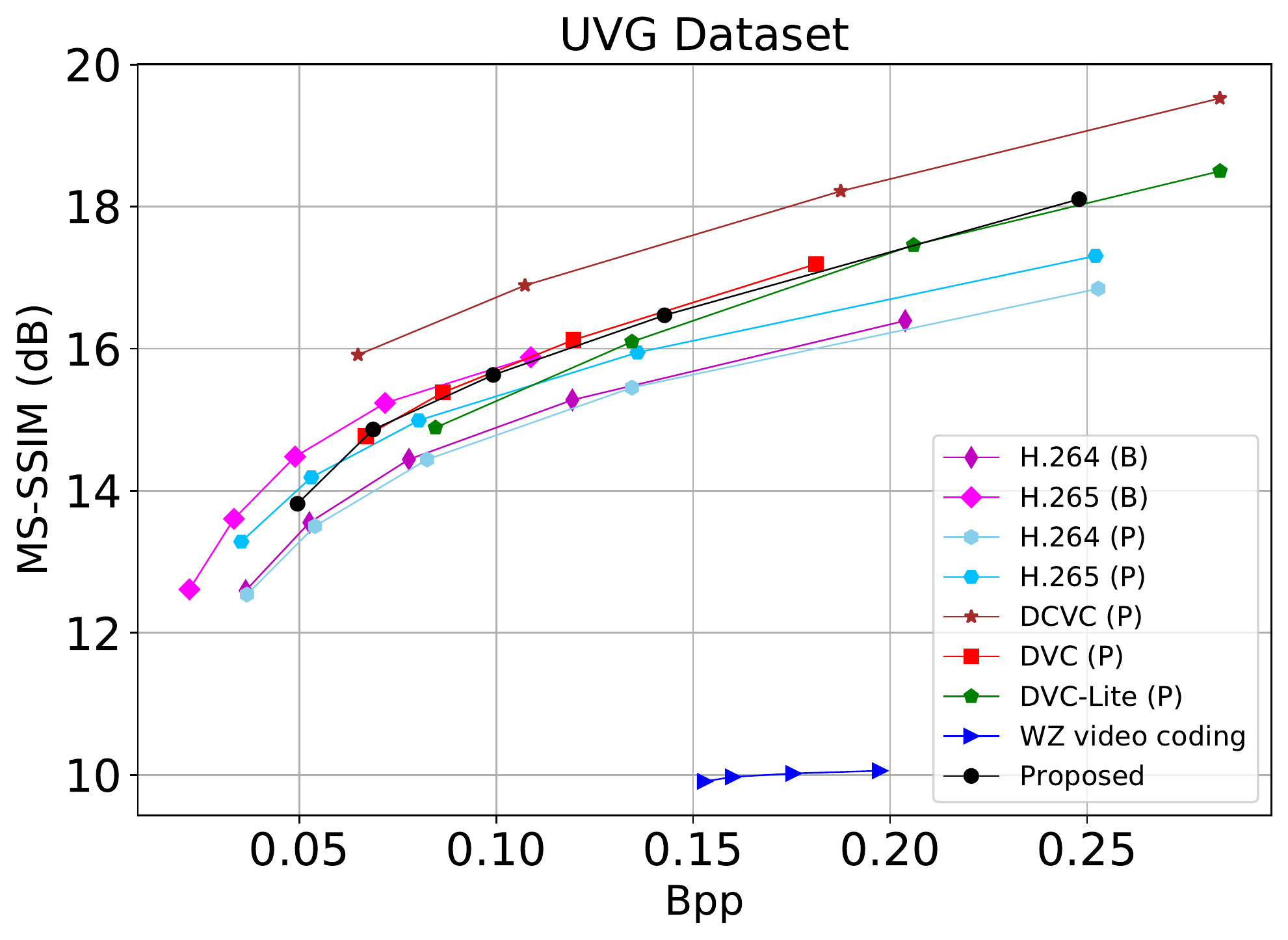}
  }
  \subfloat
  {\label{subfig:mcl_jcv_psnr}
  \includegraphics[scale=0.215]{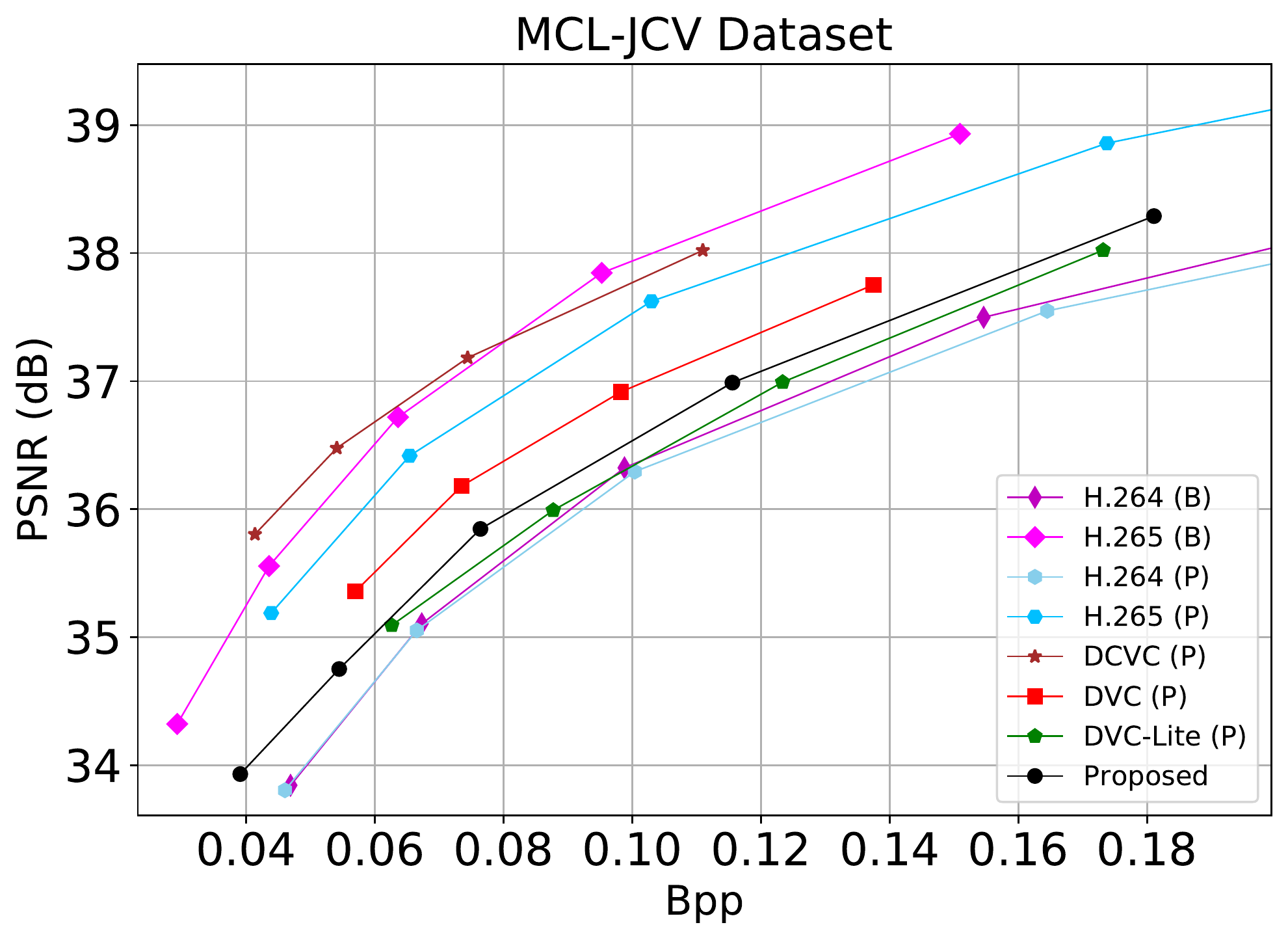}
  }
  \subfloat
  {\label{subfig:mcl_jcv_ms_ssim}
  \includegraphics[scale=0.215]{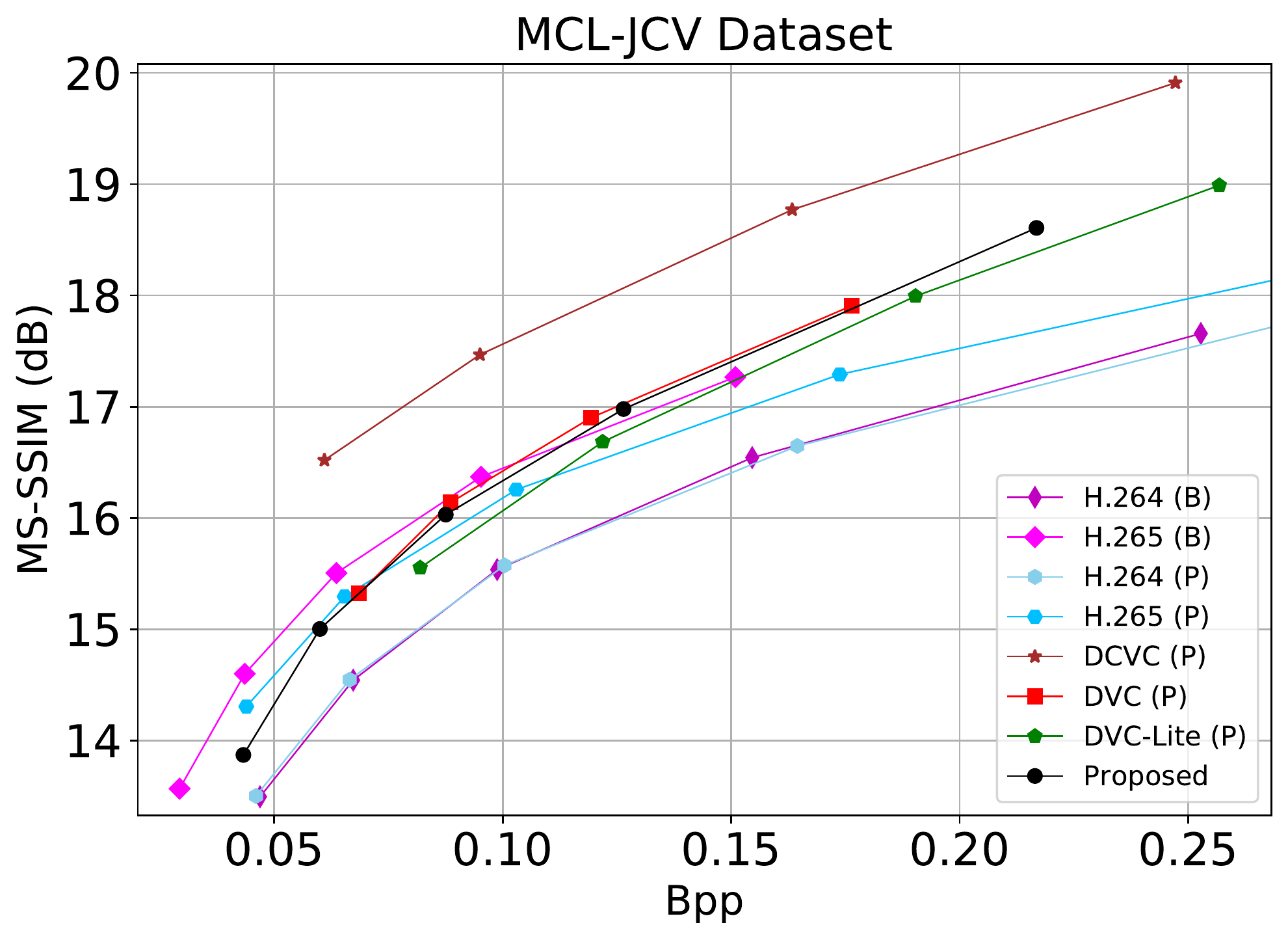}
  }
  \vspace{-0.3cm}
  \caption{Rate-distortion performance in PSNR and MS-SSIM. Due to the inefficiency of the serial implementation of the WZ video codec, we only report its results on the UVG dataset.}
  \label{fig:uvg_mcl_jcv}
  \vspace{-0.4cm}
\end{figure*}

\subsection{Training Strategy}
\label{subsec:training}
\noindent\textbf{Loss Function.} Our objective is to minimize the number of encoded bits, and reduce the distortion between the original WZ frame $x_{kN+i}$ and the reconstructed WZ frame $\hat{x}_{kN+i}$. Therefore, we use the following loss function consisting of two metrics for training:
\begin{equation}
  \begin{aligned}
    L&=\lambda D + R  \\
    &= \lambda d(x_{kN+i}, \hat{x}_{kN+i}) + R(\hat{y}_{kN+i}) + R(\hat{z}_{kN+i}) \label{eq:loss_func}
  \end{aligned}
\end{equation}
where $d(x_{kN+i}, \hat{x}_{kN+i})$ is the distortion between $x_{kN+i}$ and $\hat{x}_{kN+i}$, which can be MSE or MS-SSIM for different tasks. $R(\hat{y}_{kN+i})$ and $R(\hat{z}_{kN+i})$ denote the bits used for encoding the quantized WZ representation $\hat{y}_{kN+i}$ and the corresponding hyper representation $\hat{z}_{kN+i}$, respectively. $\lambda$ is a Lagrange multiplier that controls the trade-off between bit cost $R$ and distortion $D$. 

\noindent\textbf{Two-step Training.} During training, we adopt a two-step procedure for MSE optimized models. Firstly, we only train the WZ encoder-decoder and SI encoder with the pretrained RIFE network in \cite{huang2020rife}. After that, we jointly fine-tune the whole model including the RIFE network. For MS-SSIM optimized models, we fine-tune all of the modules based on MSE optimized models.

\section{Experiments}
\label{sec:experimets}
\subsection{Experimental Setup}
\label{subsec:experiment_setup}

\noindent\textbf{Training Data.} We use the training part of the Vimeo-90k dataset \cite{xue2019video} to train the proposed video compression framework, and randomly crop the videos into $256 \times 256$ patches. During training, we set the batch size as 16.

\noindent\textbf{Test Settings.} We evaluate the compression performance of our proposed method on two 1080p test datasets with diversified content, including UVG \cite{mercat2020uvg} and MCL-JCV \cite{wang2016mcl}. The GOP size is set as 8 by default. Our competing benchmarks contain traditional video codecs (WZ video coding \cite{kodavalla2012chroma}, H.264 \cite{wiegand2003overview} and H.265 \cite{sullivan2012overview}), and recently proposed learning-based methods (DVC \cite{lu2019dvc}, DVC-Lite \cite{lu2020end} and DCVC \cite{li2021deep}). For H.264 and H.265, we report the sequential-P order and hierarchical-B order with \textit{very slow} mode. 
For details about H.264 and H.265 settings, please refer to Appendix \ref{appendix:experiments_details}.

\noindent\textbf{Evaluation Metrics.} Since most benchmarks achieve values above 0.9 in MS-SSIM, we follow \cite{balle2018variational} to convert the quantity to decibels for improving legibility, where MS-SSIM (dB) is calculated by $10\log_{10}(1/(1-$MS-SSIM$))$. The Bjøntegaard Delta bitrate (BDBR) \cite{BDmetrics2001} is reported to denote the average bitrate savings at the same reconstruction quality. Moreover, we calculate BD-PSNR and BD-MSSSIM to indicate the average gains of reconstruction quality at the same bitrate. 

\noindent\textbf{Implementation Details.} In our experiments, we use the pretrained model \textit{mbt-2018} \cite{minnen2018joint} provided by CompressAI \cite{begaint2020compressai} for key frame compression. For WZ frame compression, we train five models with different $\lambda$ values (MSE: 0.0018, 0.0035, 0.0067, 0.0130, 0.0250; MS-SSIM: 2.40, 4.58, 8.73, 16.64, 31.73). During training, the Adam optimizer \cite{kingma2014adam} is used with an initial learning rate as 0.001. When the evaluation loss reaches a plateau, the learning rate is reduced by a factor of 2. We use a patience of 10 epochs and 5 epochs for the first and second stage, respectively. As for the total number of epochs, we set it as 100 and 50 for the two stages respectively. The whole system is implemented by PyTorch and trained on an NVIDIA RTX A5000 GPU.

\begin{table*}[t]
\footnotesize
  \caption{BDBR and BD-PSNR (BD-MSSSIM) results of learning-based video codecs when compared with H.264 (P).}
  \label{table:bd_performance}
  \vspace{-0.2cm}
  \centering
  \begin{tabular}{c|cc|cc|cc|cc}
    \hline
    \multirow{3}*{Methods} & \multicolumn{4}{c|}{UVG} & \multicolumn{4}{c}{MCL-JCV} \\
   \cline{2-9}  & \multicolumn{2}{c|}{PSNR} & \multicolumn{2}{c|}{MS-SSIM} & \multicolumn{2}{c|}{PSNR} & \multicolumn{2}{c}{MS-SSIM} \\
      & BDBR & BD-PSNR & BDBR & BD-MSSSIM & BDBR & BD-PSNR & BDBR & BD-MSSSIM \\
    \hline
    DCVC \cite{li2021deep}  & -52.96\% & 1.98dB & -59.07\% & 2.12dB & -52.18\% &  2.18dB & -61.29\% & 2.29dB \\
    DVC \cite{lu2019dvc} & -22.88\% & 0.68dB & -31.23\% &  0.75dB & -23.46\% & 0.91dB & -31.97\% &  0.83dB\\
    DVC-Lite \cite{lu2020end}   & 3.47\% & -0.14dB & -12.05\% & 0.13dB & -4.18\% & 0.55dB & -22.42\% & 0.59dB\\
    Proposed  & -17.10\% & 0.52dB & -28.82\%  & 0.64dB & -16.79\% & 0.90dB & -33.10\% &  1.00dB \\
    \hline
  \end{tabular}
  \vspace{-0.4cm}
\end{table*}

\begin{table*}[t]
\footnotesize
  \caption{Complexity of learning-based video codecs on 1080p videos. $N_c$ denotes the number of CPU cores used to encode the videos.}
  \label{table:complexity}
   \vspace{-0.2cm}
  \centering
  \begin{tabular}{c|c|ccccc|c|ccccc}
    \hline
    \multirow{2}*{Methods} & Encoding & \multicolumn{5}{c|}{Encoding Latency} & Decoding & \multicolumn{5}{c}{Decoding Latency} \\
      & FLOPs & $N_c$=1& $N_c$=2 &$N_c$=4 &$N_c$=8 & GPU & FLOPs & $N_c$=1& $N_c$=2 &$N_c$=4 &$N_c$=8 & GPU \\
    \hline
    DCVC \cite{li2021deep} & 4593.79G & 79.18s & 42.70s & 26.71s & 18.61s & 7.26s & 3030.68G & 63.38s & 45.50s & 34.30s & 27.81s & 31.80s \\
    DVC \cite{lu2019dvc} & 2859.12G & 45.16s & 24.40s & 15.51s & 9.38s & 0.44s & 1317.55G & 23.78s & 15.16s & 10.07s & 6.45s & 0.23s \\
    DVC-Lite \cite{lu2020end}  & 1062.14G & 20.69s & 12.28s & 7.82s & 4.86s & 0.31s & 932.37G & 18.31s & 10.75s & 6.82s & 4.25s & 0.21s \\
    Proposed  & 500.45G & 6.68s & 3.98s & 2.29s & 1.34s & 0.18s & 2542.98G & 22.43s & 14.09s & 9.36s & 6.49s & 0.19s \\
    \hline
  \end{tabular}
  \vspace{-0.4cm}
\end{table*}

\subsection{Experimental results}
\label{subsec:experimental_results}

\figurename~\ref{fig:uvg_mcl_jcv} shows the RD curves of different methods. It is obvious that our method outperforms the traditional WZ video codec by a large margin with $10$dB ($6.94$dB) average gains when measured by PSNR (MS-SSIM), which implies that the end-to-end optimization design can effectively improve the performance of the distributed coding. Compared with H.264 (P) and H.264 (B), our approach saves 17.10\% (16.79\%) and 12.91\% (10.85\%) bitrates in terms of PSNR on UVG (MCL-JCV) dataset, respectively. In comparison with H.265, our method achieves better performance under the metric of MS-SSIM except for very low-rate regime. \tableautorefname~\ref{table:bd_performance} reports the BDBR and BD-PSNR (BD-MSSSIM) results of different video compression methods when compared with H.264 (P) on UVG and MCL-JCV datasets. We see that our method achieves coding gains comparable to DVC and surpasses DVC-Lite, which demonstrate that our Distributed DVC is a feasible attempt to approach the performance of predictive coding. More results for per video are given in Appendix \ref{appendix:rd_performance}. 

We also compare the complexity of four learning-based video codecs on 1080p videos. As shown in \tableautorefname~\ref{table:complexity}, on CPU platforms with different computing powers (i.e., $N_c$=1,2,4,8), our method achieves 3$\sim$3.6$\times$ encoding speedup against DVC-Lite and saves more bits. When compared with DVC, our method achieves 6$\sim$7$\times$ encoding speedup and similar decoding latency, only dropping 0.033dB and 0.049dB in PSNR on UVG and MCL-JCV dataset, respectively. When the encoder is implemented on a powerful GPU, our method still saves 60\% of the encoding time. Although there is a performance gap between the proposed method and DCVC, our method can reduce about 90\% of the encoding latency. These results demonstrate that our video codec based on a distributed coding architecture has the potential to satisfy the needs of uplink-based applications requiring low complexity encoders.



\begin{figure}[t]
  \centering
  \includegraphics[scale=0.25]{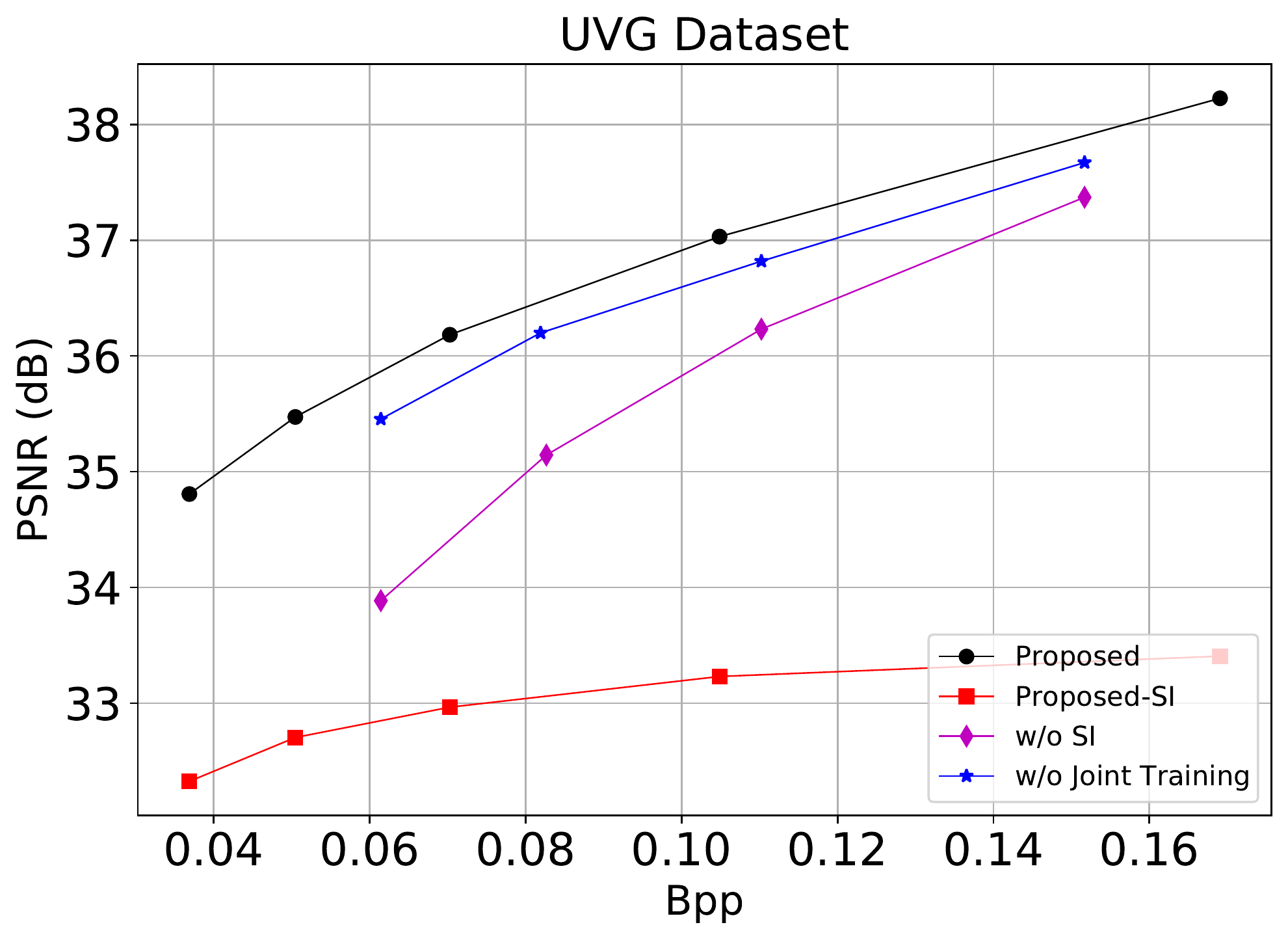}
  \vspace{-0.3cm}
  \caption{Ablation study. \textit{Proposed-SI} means the quality of SI frame. \textit{w/o SI} denotes removing the SI generation network at the decoder. \textit{w/o Joint Training} is to fix the pretrained encoder including the entropy model and only train the decoder part.}
  \label{fig:ablation_training}
  \vspace{-0.5cm}
  \end{figure}
\subsection{Ablation Study}

\noindent\textbf{Side Information.}
As mentioned in Sec.~\ref{subsec:side_information}, we propose an SI generation network to exploit temporal correlations. To verify its effectiveness, we perform a set of ablation experiments on the UVG dataset. As shown in \figurename~\ref{fig:ablation_training}, we provide qualities of the SI frame under different bpp levels. After removing the SI generation network, the PSNR of \textit{w/o SI} drops by about 1.17 dB at the same bpp level, which illustrates that the SI frame helps our WZ decoder reconstruct the higher quality frame. In \tableautorefname~\ref{table:si_generation}, we further analyse each component in the SI generation network. Fixing the RIFE network leads to 8.05\% increase in bitrate and 0.18dB decrease in PSNR, because the pretrained RIFE network only aims at for estimating the intermediate frame more accurately, but not for optimizing the RD performance. In addition, if we do not generate the intermediate frame and concatenate two decoded frames following the hierarchical order as the SI, it requires 25.68\% more bitrate and drops PSNR by 0.54dB, highlighting the benefit of video interpolation. Moreover, the SI exploited in the pixel space results in up to 48.86\% bitrate consumption and 0.94dB gain decrease, supporting the necessity of processing SI in the feature space.

\noindent\textbf{Joint Encoder-Decoder Training Strategy.} 
In this paper, we jointly optimize the encoder and decoder to implicitly reduce the partial temporal redundancy between frames, which makes the latent representation $\hat{y}_{kN+i}$ more compact. By only training the decoder networks with the fixed pre-trained encoder and entropy model, \textit{w/o Joint Training} method shown in \figurename~\ref{fig:ablation_training} consumes more bits, which indicates that the latent representation of our method have more elements with low magnitudes requiring much fewer bits for encoding.

\section{Conclusions and Discussions}
\label{sec:conclusions}
In this paper, we present an end-to-end distributed deep video compression framework to improve the compression performance of distributed video coding. Our proposal inherits the merits of traditional distributed video coding in the low encoding complexity and learning-based compression in the powerful non-linear representation ability. Experimental results show the competence of our framework in achieving a better coding efficiency than traditional distributed video coding methods and H.264. Moreover, when compared with DVC, our proposed method enjoys a much lower encoder complexity with competitive coding performance. Overall, our framework is promising in enabling deep video compression systems with low-complexity encoders. 

Although there is still a performance gap to deep video compression with the predictive coding architecture, we believe it can be narrowed by leveraging the latest advancements in deep learning to further improve the coding efficiency. For example, scale flow \cite{agustsson2020scale} and deformable compensation \cite{hu2021fvc} can be applied to improve the SI quality. Moreover, our framework enjoys great potentials in application scenarios with multiple video sources captured by different camera sensors \cite{zhang2023ldmic}, where only the decoder can exploit the inter-view redundancy. Therefore, it deserves more research efforts to explore the potential of our framework.


\begin{table}[t]
\footnotesize
  \caption{Ablation study on side information component. The first row denotes our final solution and is selected as the anchor.}
  \label{table:si_generation}
  \vspace{-0.2cm}
  \centering
  \begin{tabular}{ccc|cc}
    \hline
    \multirow{1}*{RIFE} & \multirow{1}*{SI}& \multirow{1}*{Joint} & \multirow{1}*{Bitrate} & \multirow{1}*{PSNR} \\
    Network & Encoder & Fine-tuning & Increase & Decrease \\
    \hline
    \checkmark & \checkmark & \checkmark & 0\% & 0dB  \\
    \checkmark & \checkmark &  & 8.05\% & 0.18dB \\
     & \checkmark & \checkmark & 25.68\% & 0.54dB\\
    \checkmark &  & \checkmark & 48.86\% & 0.94dB\\
    \hline
  \end{tabular}
  \vspace{-0.5cm}
\end{table}
\section*{Acknowledgment}
This work was supported by the General Research Fund (Project No. 16209622) from the Hong Kong Research Grants Council.

\vspace{-0.2cm}

\bibliographystyle{ieeetr} 
\bibliography{icme_references}

\begin{thebibliography}{10}

\bibitem{lu2019dvc}
G.~Lu, W.~Ouyang, D.~Xu, X.~Zhang, C.~Cai, and Z.~Gao, ``Dvc: An end-to-end
  deep video compression framework,'' in {\em CVPR}, 2019.

\bibitem{lu2020end}
G.~Lu, X.~Zhang, W.~Ouyang, L.~Chen, Z.~Gao, and D.~Xu, ``An end-to-end
  learning framework for video compression,'' {\em TPAMI}, vol.~43, no.~10,
  pp.~3292--3308, 2020.

\bibitem{agustsson2020scale}
E.~Agustsson, D.~Minnen, N.~Johnston, J.~Balle, S.~J. Hwang, and G.~Toderici,
  ``Scale-space flow for end-to-end optimized video compression,'' in {\em
  CVPR}, 2020.

\bibitem{hu2021fvc}
Z.~Hu, G.~Lu, and D.~Xu, ``Fvc: A new framework towards deep video compression
  in feature space,'' in {\em CVPR}, 2021.

\bibitem{li2021deep}
J.~Li, B.~Li, and Y.~Lu, ``Deep contextual video compression,'' in {\em
  NeurIPS}, 2021.

\bibitem{li2022hybrid}
J.~Li, B.~Li, and Y.~Lu, ``Hybrid spatial-temporal entropy modelling for neural
  video compression,'' in {\em ACM MM}, 2022.

\bibitem{wiegand2003overview}
T.~Wiegand, G.~J. Sullivan, G.~Bjontegaard, and A.~Luthra, ``Overview of the h.
  264/avc video coding standard,'' {\em TCSVT}, vol.~13, no.~7, pp.~560--576,
  2003.

\bibitem{sullivan2012overview}
G.~J. Sullivan, J.-R. Ohm, W.-J. Han, and T.~Wiegand, ``Overview of the high
  efficiency video coding (hevc) standard,'' {\em TCSVT}, vol.~22, no.~12,
  pp.~1649--1668, 2012.

\bibitem{fan2019survey}
C.-L. Fan, W.-C. Lo, Y.-T. Pai, and C.-H. Hsu, ``A survey on 360 video
  streaming: Acquisition, transmission, and display,'' {\em CSUR}, vol.~52,
  no.~4, pp.~1--36, 2019.

\bibitem{miyagawa2014overview}
N.~Miyagawa, ``Overview of blu-ray disc™ recordable/rewritable media
  technology,'' {\em FOE}, vol.~7, no.~4, pp.~409--424, 2014.

\bibitem{elharrouss2021review}
O.~Elharrouss, N.~Almaadeed, and S.~Al-Maadeed, ``A review of video
  surveillance systems,'' {\em JVCIR}, vol.~77, p.~103116, 2021.

\bibitem{zhang2023ldmic}
X.~Zhang, J.~Shao, and J.~Zhang, ``Ldmic: Learning-based distributed multi-view
  image coding,'' in {\em ICLR}, 2023.

\bibitem{Slepian1973}
D.~Slepian and J.~Wolf, ``Noiseless coding of correlated information sources,''
  {\em TIT}, vol.~19, no.~4, pp.~471--480, 1973.

\bibitem{Wyner1976}
A.~Wyner and J.~Ziv, ``The rate-distortion function for source coding with side
  information at the decoder,'' {\em TIT}, vol.~22, no.~1, pp.~1--10, 1976.

\bibitem{Girod2005}
B.~Girod, A.~M. Aaron, S.~Rane, and D.~Rebollo-Monedero, ``Distributed video
  coding,'' {\em Proceedings of the IEEE}, vol.~93, no.~1, pp.~71--83, 2005.

\bibitem{kodavalla2011chroma}
V.~K. Kodavalla and P.~K. Mohan, ``Chroma components coding in feedback-free
  distributed video coding,'' in {\em IMSAA}, 2011.

\bibitem{kodavalla2012chroma}
V.~K. Kodavalla and P.~K. Mohan, ``Chroma components coding method in
  distributed video coding,'' in {\em ICDCS}, 2012.

\bibitem{dufaux2010distributed}
F.~Dufaux, W.~Gao, S.~Tubaro, and A.~Vetro, ``Distributed video coding: trends
  and perspectives,'' {\em JIVP}, vol.~2009, pp.~1--13, 2010.

\bibitem{aaron2002wyner}
A.~Aaron, R.~Zhang, and B.~Girod, ``Wyner-ziv coding of motion video,'' in {\em
  ASILOMAR}, 2002.

\bibitem{dash2018decoder}
B.~Dash, S.~Rup, A.~Mohapatra, B.~Majhi, and M.~Swamy, ``Decoder driven side
  information generation using ensemble of mlp networks for distributed video
  coding,'' {\em Multimed. Tools. Appl.}, vol.~77, no.~12, pp.~15221--15250,
  2018.

\bibitem{dash2019decoder}
B.~Dash, S.~Rup, A.~Mohapatra, B.~Majhi, and M.~Swamy, ``Decoder side
  wyner--ziv frame estimation using chebyshev polynomial-based flann technique
  for distributed video coding,'' {\em Multidim. Syst. Sign. P.}, vol.~30,
  no.~3, pp.~1031--1061, 2019.

\bibitem{minnen2020channel}
D.~Minnen and S.~Singh, ``Channel-wise autoregressive entropy models for
  learned image compression,'' in {\em ICIP}, 2020.

\bibitem{balle2017end}
J.~Ball{\'e}, V.~Laparra, and E.~P. Simoncelli, ``End-to-end optimized image
  compression,'' in {\em ICLR}, 2017.

\bibitem{balle2018variational}
J.~Ball{\'e}, D.~Minnen, S.~Singh, S.~J. Hwang, and N.~Johnston, ``Variational
  image compression with a scale hyperprior,'' in {\em ICLR}, 2018.

\bibitem{minnen2018joint}
D.~Minnen, J.~Ball{\'e}, and G.~D. Toderici, ``Joint autoregressive and
  hierarchical priors for learned image compression,'' in {\em NeurIPS}, 2018.

\bibitem{cheng2020learned}
Z.~Cheng, H.~Sun, M.~Takeuchi, and J.~Katto, ``Learned image compression with
  discretized gaussian mixture likelihoods and attention modules,'' in {\em
  CVPR}, 2020.

\bibitem{he2021checkerboard}
D.~He, Y.~Zheng, B.~Sun, Y.~Wang, and H.~Qin, ``Checkerboard context model for
  efficient learned image compression,'' in {\em CVPR}, 2021.

\bibitem{balle2016density}
J.~Ball{\'e}, V.~Laparra, and E.~P. Simoncelli, ``Density modeling of images
  using a generalized normalization transformation,'' in {\em ICLR}, 2016.

\bibitem{hong2020efficient}
W.~Hong, T.~Chen, M.~Lu, S.~Pu, and Z.~Ma, ``Efficient neural image decoding
  via fixed-point inference,'' {\em TCSVT}, vol.~31, no.~9, pp.~3618--3630,
  2020.

\bibitem{sun2021end}
H.~Sun, L.~Yu, and J.~Katto, ``End-to-end learned image compression with
  quantized weights and activations,'' {\em arXiv}, 2021.

\bibitem{jia2022fpx}
C.~Jia, X.~Hang, S.~Wang, Y.~Wu, S.~Ma, and W.~Gao, ``Fpx-nic: An
  fpga-accelerated 4k ultra-high-definition neural video coding system,'' {\em
  TCSVT}, vol.~32, no.~9, pp.~6385--6399, 2022.

\bibitem{aaron2004transform}
A.~Aaron, S.~D. Rane, E.~Setton, and B.~Girod, ``Transform-domain wyner-ziv
  codec for video,'' in {\em VCIP}, 2004.

\bibitem{brites2008correlation}
C.~Brites and F.~Pereira, ``Correlation noise modeling for efficient pixel and
  transform domain wyner--ziv video coding,'' {\em TCSVT}, vol.~18, no.~9,
  pp.~1177--1190, 2008.

\bibitem{artigas2007discover}
X.~Artigas, J.~Ascenso, M.~Dalai, S.~Klomp, D.~Kubasov, and M.~Ouaret, ``The
  discover codec: architecture, techniques and evaluation,'' in {\em PCS},
  2007.

\bibitem{ren2011new}
P.~Ren, P.~Shi, C.~Luo, and Q.~Liu, ``A new scheme for side information
  generation in dvc by using optical flow algorithm,'' in {\em ICMT}, 2011.

\bibitem{bhagath2016low}
P.~R. Bhagath, J.~Mukherjee, and S.~Mukopadhayay, ``Low complexity encoder for
  feedback-channel-free distributed video coding using deep convolutional
  neural networks at the decoder,'' in {\em ICVGIP}, 2016.

\bibitem{duda2013asymmetric}
J.~Duda, ``Asymmetric numeral systems: entropy coding combining speed of
  huffman coding with compression rate of arithmetic coding,'' {\em arXiv},
  2013.

\bibitem{huang2020rife}
Z.~Huang, T.~Zhang, W.~Heng, B.~Shi, and S.~Zhou, ``Rife: Real-time
  intermediate flow estimation for video frame interpolation,'' in {\em ECCV},
  2022.

\bibitem{xue2019video}
T.~Xue, B.~Chen, J.~Wu, D.~Wei, and W.~T. Freeman, ``Video enhancement with
  task-oriented flow,'' {\em IJCV}, vol.~127, no.~8, pp.~1106--1125, 2019.

\bibitem{mercat2020uvg}
A.~Mercat, M.~Viitanen, and J.~Vanne, ``Uvg dataset: 50/120fps 4k sequences for
  video codec analysis and development,'' in {\em MMSys}, 2020.

\bibitem{wang2016mcl}
H.~Wang, W.~Gan, S.~Hu, J.~Y. Lin, L.~Jin, L.~Song, P.~Wang, I.~Katsavounidis,
  A.~Aaron, and C.-C.~J. Kuo, ``Mcl-jcv: a jnd-based h. 264/avc video quality
  assessment dataset,'' in {\em ICIP}, 2016.

\bibitem{BDmetrics2001}
G.~Bjøntegaard, ``Calculation of average psnr differences between rd-curves,''
  {\em ITU-T VCEG-M33}, 2001.

\bibitem{begaint2020compressai}
J.~B{\'e}gaint, F.~Racap{\'e}, S.~Feltman, and A.~Pushparaja, ``Compressai: a
  pytorch library and evaluation platform for end-to-end compression
  research,'' {\em arXiv}, 2020.

\bibitem{kingma2014adam}
D.~P. Kingma and J.~Ba, ``Adam: A method for stochastic optimization,'' in {\em
  ICLR}, 2015.

\end{thebibliography}

\newpage

\appendix

\section{Distributed Coding}
\subsection{Foundation}
\label{appendix:sw_and_wz_theorem}

\begin{figure*}[t]
  \centering
  \subfloat[] 
  {\label{subfig:SW_arch}
  \includegraphics[scale=0.45]{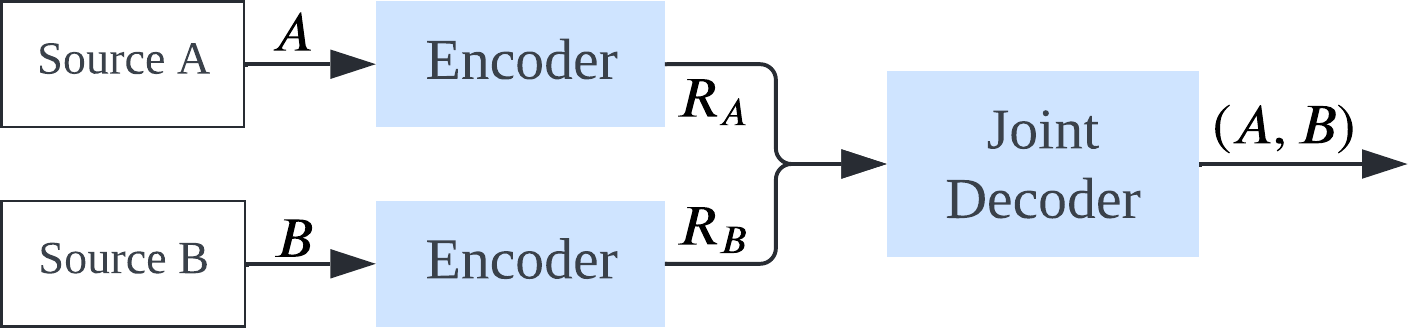}
  }
  \subfloat[]
  {\label{subfig:WZ_arch}
  \includegraphics[scale=0.45]{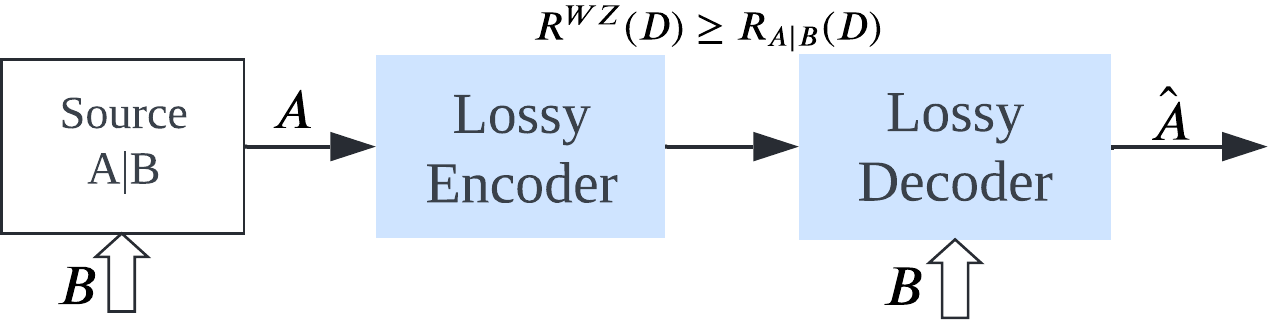}
  }
  \caption{(a): Distributed compression of two correlated sources $A$ and $B$. The decoder jointly decompress $A$ and $B$ to utilize their mutual dependence. (b): Lossy compression of a source $A$ using statistically related side information $B$.}
  \label{fig:SW_WZ_arch}
\end{figure*}

\begin{figure*}[t]
\centering
\includegraphics[scale=0.3]{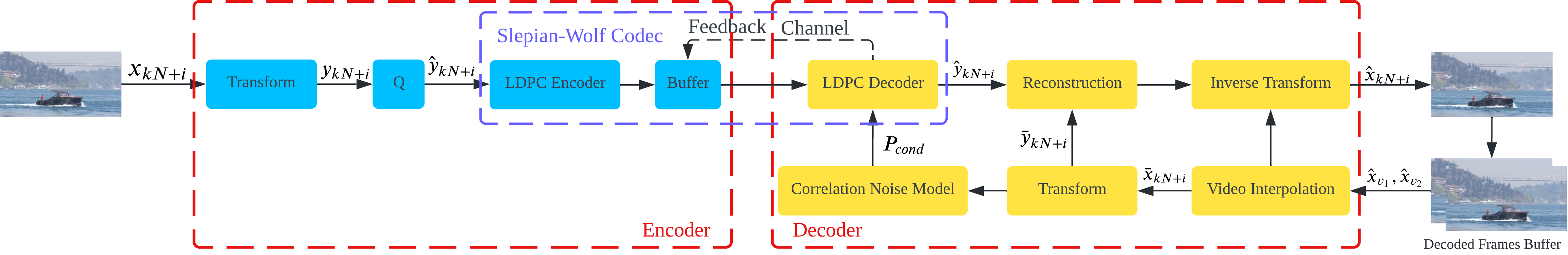}
\caption{The traditional WZ video coding architecture.}
\label{fig:WZ_Video_Coding}
\vspace{-0.4cm}
\end{figure*}

As shown in \figurename~\ref{fig:SW_WZ_arch} (a), distributed coding refers to separate encoding and joint decoding for two (or more) statistically correlated but physically separated sources. The joint decoding procedure aims at exploiting statistical correlations across different sources to achieve efficient compression.
Slepian and Wolf studied the distributed lossless coding problem in the basic two-source case in 1973.
Wyner and Ziv then extended the Slepian and Wolf (SW) theorem to the lossy case, namely the Wyner–Ziv (WZ) theorem, which presents the achievable lower bound for the bit rate at given distortion, as shown in \figurename~\ref{fig:SW_WZ_arch} (b). This architecture has been extended to the video coding area, called as distributed video coding \cite{Girod2005}, which aims at independent encoding of each frame and joint decoding with side information generated from previously decoded frames.
The formal statements of SW theorem and WZ theorem are as follows.

\begin{theorem}[Slepian-Wolf]
Consider two statistically dependent i.i.d. sources A and B, the achievable rate region of compressing A and B without any distortion, is provided by:
$$R_A \geq H(A|B),R_B \geq H(B|A),R_A+R_B\geq H(A)+H(B),$$
where $R_A$ and $R_B$ represent the rates for transmitting $A$ and $B$, respectively.
\end{theorem}

\begin{theorem}[Wyner-Ziv]
Assume sources A and B are statisctially correlated. 
Given a certain distortion level $D$, the minimum rate $R_{WZ}(D)$ for encoding $A$ with side information $B$ at the decoder is larger than or equal to $R_{A|B}(D)$, where $B$ is available at both the encoder and the decoder.
Denoting the output of decoder as $\hat{A}$, we have the following equivalence: $$R_{WZ}(D)\geq R_{A|B}(D)=\min_{E[d(A,\hat{A})]\leq D} I(A;\hat{A}|B),$$
where $R_{WZ}(D)=R_{A|B}(D)$ when A and B are jointly Gaussian, and the distortion $d(A,\hat{A})$ between ${A}$ and $\hat{A}$ is measured by the mean-squared error.
\end{theorem}

The SW theorem proclaims that lossless compression of two correlated data sources with separate encoders and a joint decoder can asymptotically achieve the same compression rate as the optimal compression with a joint encoder and decoder. The WZ theorem extends this idea to lossy compression and demonstrates that there is no RD performance loss without the side information at the encoder. 
The SW and WZ theorems imply that it is possible to compress two statistically dependent signals in a distributed way (separate encoding and joint decoding) while achieving the RD performance of predictive coding methods (joint encoding and decoding). For more details on distributed coding and its applications in video coding, please refer to \cite{Girod2005} and \cite{dufaux2010distributed}.

\subsection{Brief Introduction of Classic WZ Video Coding}
\label{appendix:classic_wz_video_coding}
\figurename~\ref{fig:WZ_Video_Coding} illustrates the classic architecture of WZ video coding \cite{kodavalla2012chroma}. The encoding and decoding procedure of the WZ video compression is briefly summarized as follows,

\textbf{Step 1. Transformation and quantization.} 
The input WZ frame $x_{kN+i}$ is transformed to $y_{kN+i}$ by applying discrete cosine transform (DCT). Then $y_{kN+i}$ is uniformly quantized to $\hat{y}_{kN+i}$ according to predefined quantization levels $\mathcal{Q}=\{0, 2^m|m=1,...,7\}$. The value 0 indicates that some transform bands are not encoded and is replaced by the SI's corresponding bands at the decoder, while the other bands are divided into multiple bit planes that are processed by next module.

\textbf{Step 2. Slepian-Wolf encoding.}
An LDPC accumulate encoder is used to encode the bit planes of $\hat{y}_{kN+i}$ separately and generate the corresponding parity information to be stored in a buffer and sent in chunks upon the request from the decoder via the feedback channel.

\textbf{Step 3. Side information generation.}
Based on two previous decoded frames $\hat{x}_{v_1}, \hat{x}_{v_2}$ and the hierarchical frame interpolation order shown in \figurename~\ref{fig:interpolation_order}, a motion compensated frame interpolation algorithm is used to create an SI frame $\bar{x}_{kN+i}$ that is transformed to $\bar{y}_{kN+i}$. The correlation noise between $\hat{y}_{kN+i}$ and $\bar{y}_{kN+i}$ is modeled by a Laplacian distribution as a virtual channel model. Then the soft information (\textit{i.e.}, conditional bit probabilities $P_{cond}$) for each bitplane is estimated by using the SI representation $\bar{y}_{kN+i}$, the correlation noise and the previous decoded bit planes.

\textbf{Step 4. Slepian-Wolf decoding.}
Given the soft information, each bit plane is decoded by requesting the successive chunks of parity bits from the encoder buffer through the feedback channel until a low bit error probability is achieved.


\textbf{Step 5. Reconstruction and inverse transformation.}
The transform bands are firstly reconstructed by grouping the SI's high frequency bands and the decoded bands, followed by de-quantization and inverse transform to obtain the reconstructed WZ frame $\hat{x}_{kN+i}$.

\subsection{Network Architecture}
\label{appendix:network_arch}

\begin{figure}[t]
  \centering
  \includegraphics[scale=0.4]{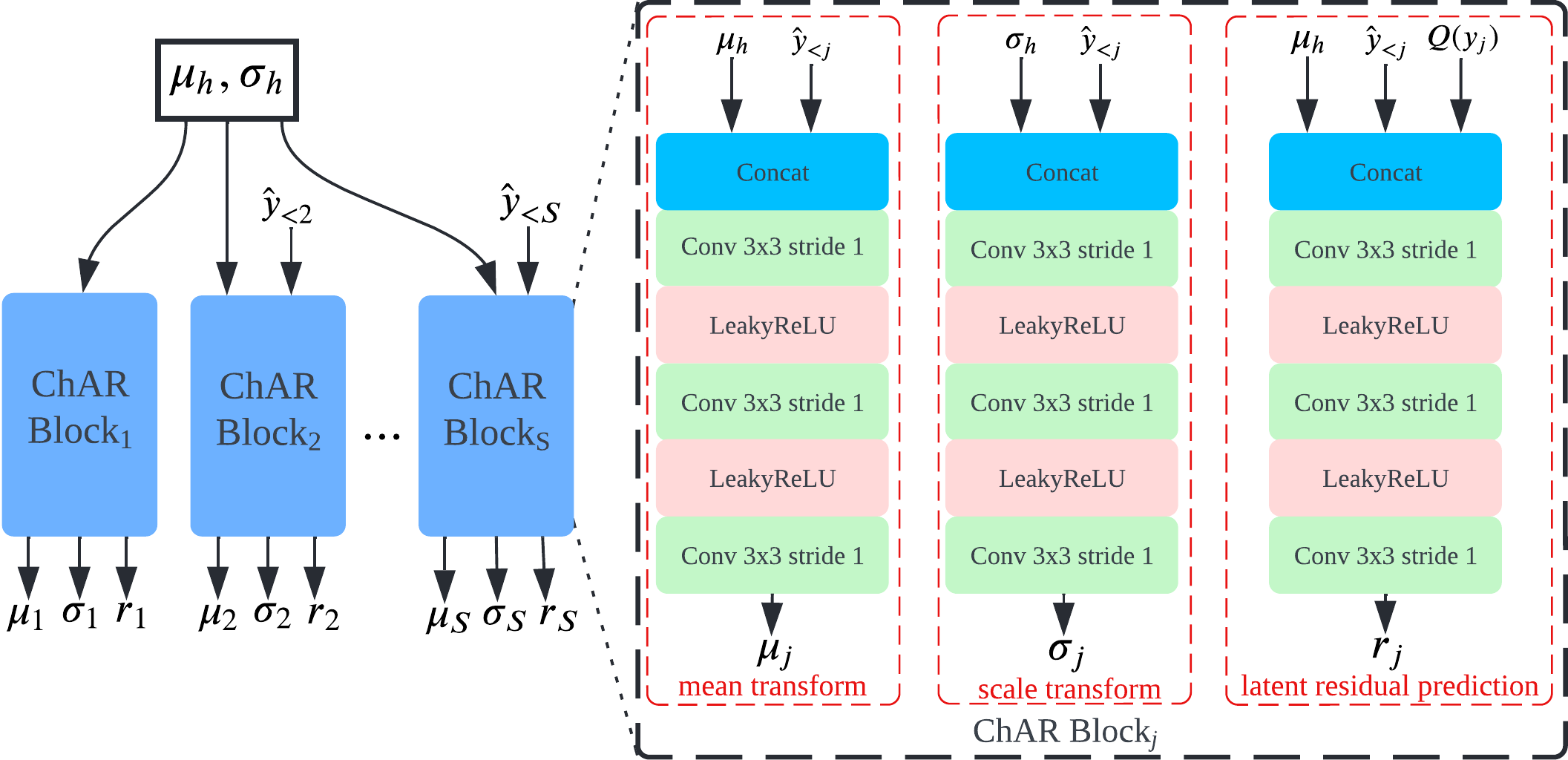}
  \caption{Network structure of ChAR component. $\mu_h$ and $\sigma_h$ represent the estimated mean and variance from the hyperprior entropy model. The $S$ ChAR blocks are performed sequentially since $\hat{y}_{j}$ is decoded after $\mu_j$ and $\sigma_j$ is obtained. We set $S$ as $8$ in the proposed model.}
  \label{fig:ChAR}
\end{figure}

\begin{figure*}[t]
  \centering
  \subfloat[Network structure of IFNet. Each IFBlock has a resolution parameter, i.e., $(K_0, K_1, K_2) = (4, 2, 1)$.] 
  {\label{subfig:IFNet}
  \includegraphics[scale=0.4]{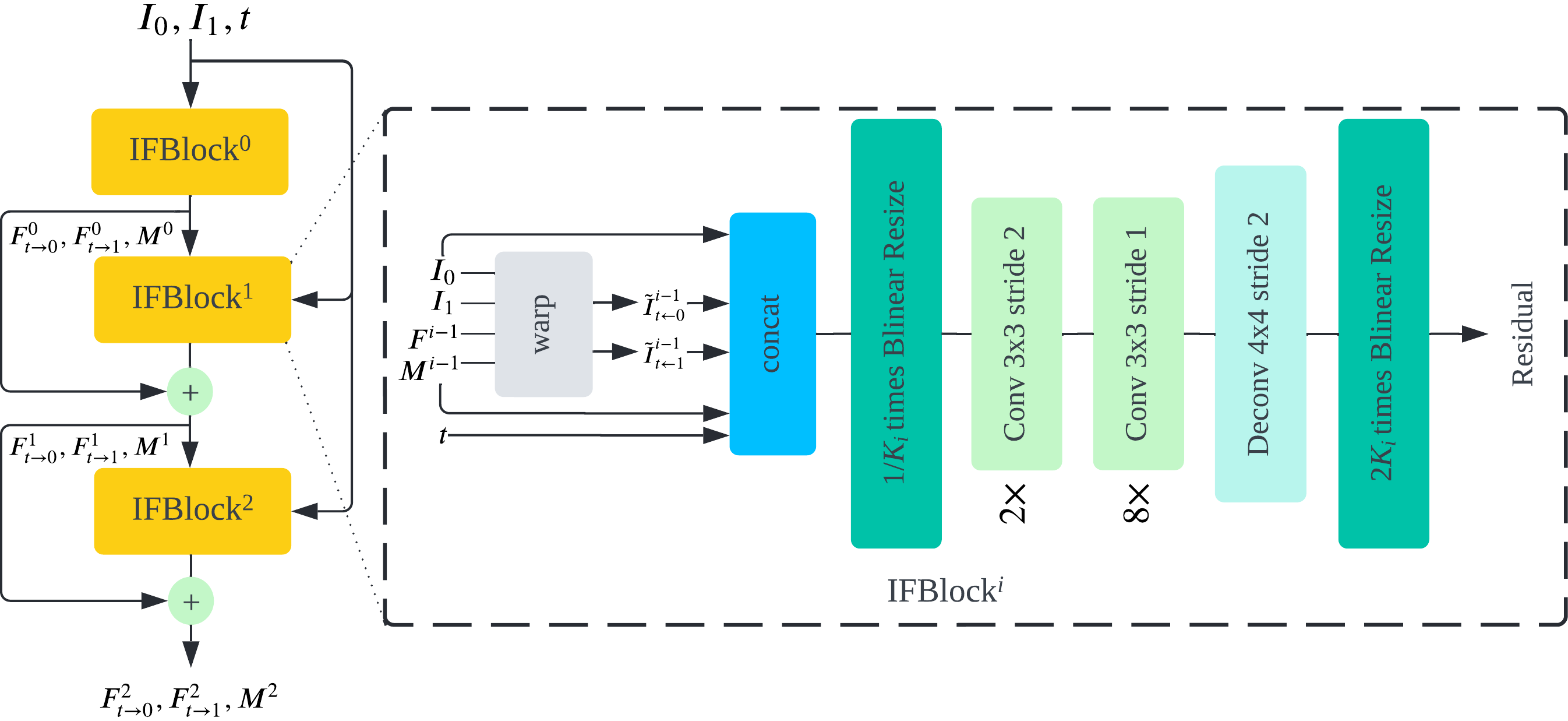}
  } \\
  \subfloat[Network structure of RefineNet with a context extractor and an Unet refine network.]
  {\label{subfig:RefineNet}
  \includegraphics[scale=0.35]{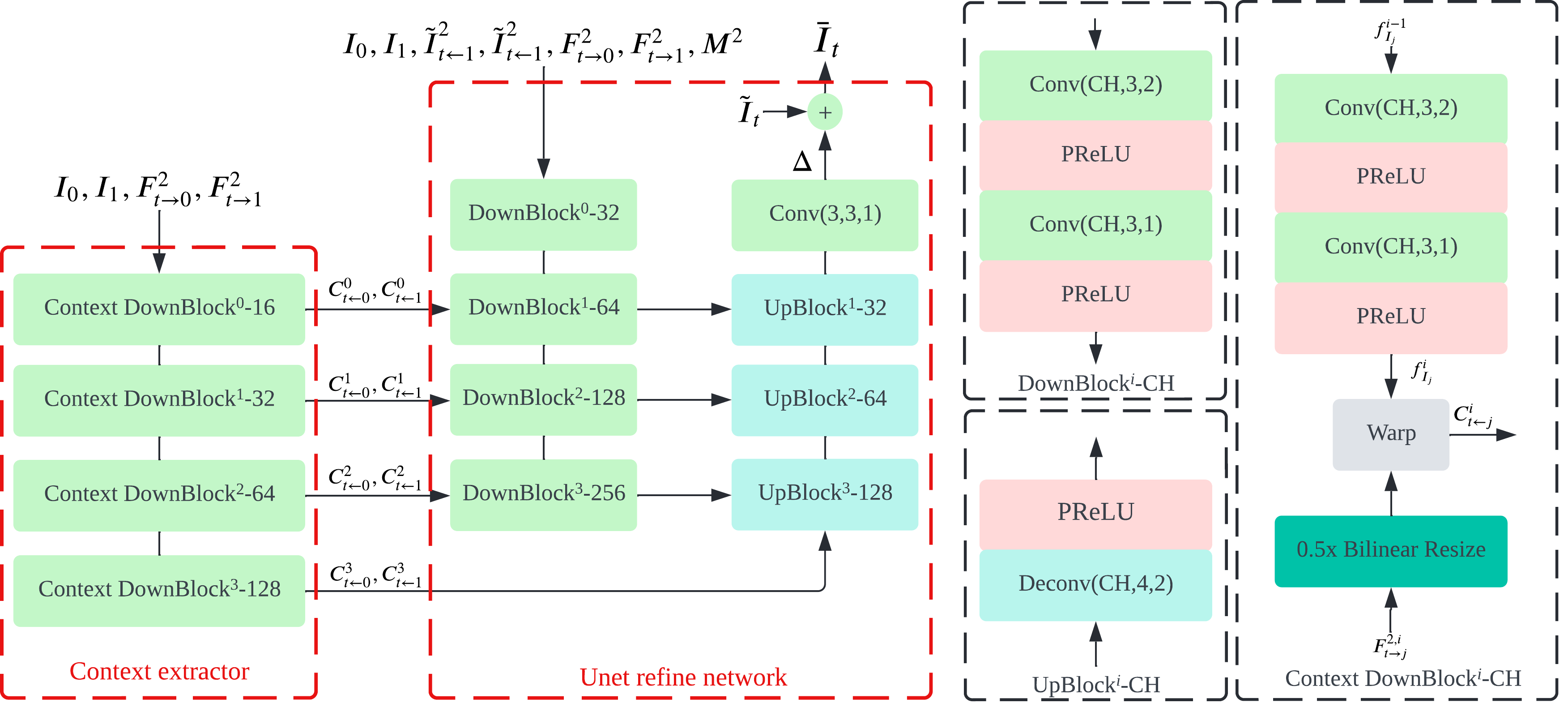}
  }
  \caption{Network structure of RIFE.}
  \label{fig:RIFE_network}
  \vspace{-0.4cm}
\end{figure*}

\textbf{ChAR component.} As shown in \figurename~\ref{fig:ChAR}, each ChAR block is composed of three modules, including a mean transform module, a scale transform module, and a latent residual prediction module. It is noted that the quantized slice $Q(y_j)=Round(y_j-\mu_j)+\mu_j$ is concatenated with the current slice's mean $\mu_j$ and the previous decoded slices $\hat{y}_{<j}$ to obtain the predicted residual $r_j$. To reduce the quantization error, we sum $r_j$ and $Q(y_j)$ together to obtain the current decoded slice $\hat{y}_{j}$. 

\noindent\textbf{RIFE network.} 
As shown in \figurename~\ref{fig:RIFE_network}, the RIFE network is composed of an intermediate flow estimation network (IFNet) and a RefineNet. Given two previous decoded frames $\hat{x}_{v_1}, \hat{x}_{v_2}$ and the corresponding time step $t$ $(0\leq t \leq 1)$, the IFNet estimates the motion information and produces a coarse interpolated frame. Then the RefineNet is used to refine the high-frequency area and reduce artifacts to create the current predicted frame $\bar{x}_{kN+i}$, which is expected to be as close to the current frame $x_{kN+i}$ as possible. The IFNet adopts several stacked IFBlocks at different resolution to obtain a rough interpolated frame with the following formula:
\begin{equation}
\begin{gathered}
\tilde{I}_{t}=M \odot \tilde{I}_{t \leftarrow 0}+(1-M) \odot \tilde{I}_{t \leftarrow 1} \\
\tilde{I}_{t \leftarrow 0}=\overleftarrow{\mathcal{W}}\left(I_{0}, F_{t \rightarrow 0}\right), \quad \tilde{I}_{t \leftarrow 1}=\overleftarrow{\mathcal{W}}\left(I_{1}, F_{t \rightarrow 1}\right) \label{eq:rough_frame}
\end{gathered}
\end{equation}
where $M$ denotes the fusion map $(0 \leq M \leq 1)$, the operation $\odot$ is an element-wise multiplier, and $\overleftarrow{\mathcal{W}}$ represents the image backward warping. 
For each IFBlock, two input frames $I_0$ and $I_1$ are first warped to the current frames $\tilde{I}_{t\leftarrow0}$ and $\tilde{I}_{t\leftarrow1}$  based on estimated flow $F^{i-1}$ from the $(i-1)^{th}$ IFBlock. Then we concatenate the input frames $I_0, I_1$, warped frames $\tilde{I}_{t\leftarrow0}^{i-1},\tilde{I}_{t\leftarrow1}^{i-1}$, current timestep $t$, previous flow $F^{i-1}$ and fusion map $M^{i-1}$ by channel dimension, and feed them into a series of bilinear and convolution operations to approximate the residual of flow and fusion map.
After obtaining the the final flow $F^{2}$ and the fusion map $M^{2}$, we use equation \equationautorefname~\ref{eq:rough_frame} to get the interpolated frame $\tilde{I}_{t}$.

To refine the high frequency area and reduce the artifacts of $\tilde{I}_{t}$, the RefineNet is employed to produce a reconstruction residual $\Delta (-1 \leq \Delta \leq 1)$. Specifically, the context extractor first extracts the multi-scale contextual features $C_0$ and $C_1$ from input frames $I_0$ and $I_1$, respectively.
Based on the intermediate flows $F_{t \rightarrow 0}^2$ and $F_{t \rightarrow 1}^2$, these features are warped to the aligned features $C_{t \leftarrow 0}$ and $C_{t \leftarrow 1}$. 
At the same time, the input frames $I_0, I_1$, warped frames $\tilde{I}_{t\leftarrow0}^{2},\tilde{I}_{t\leftarrow1}^{2}$, the estimated flows $F_{t \rightarrow 0}^2, F_{t \rightarrow 1}^2$ and the fusion map $M^{2}$ are concatenated and fed into the encoder of the Unet refine network to produce a refined reconstructed frame $\bar{I}_t=\tilde{I}_{t}+\Delta$ with the aid of $C_{t\leftarrow0}$ and $C_{t\leftarrow1}$.

\section{Experiments}
\subsection{Experimental Details}
\label{appendix:experiments_details}

\textbf{Key frame compression.}
In learning-based video codecs, the pretrained model \textit{mbt-2018} \cite{minnen2018joint} with quality index 4 provided by CompressAI \cite{begaint2020compressai} is used to compress key frames (\textit{i.e.}, the first frame of each GOP). 
For fair comparison, both DVC  \footnote{https://github.com/ZhihaoHu/PyTorchVideoCompression/tree/master/DVC} \cite{lu2019dvc} and DCVC \footnote{https://github.com/DeepMC-DCVC/DCVC} \cite{li2021deep} with the above intra-coding method are retested by using their open source codes.
For traditional video codecs, H.264 and H.265 utilize their default intra-coding methods, and WZ video coding \cite{kodavalla2012chroma} adopt H.265 intra coding method to compress key frames.

\noindent\textbf{H.264 and H.265 settings.}
We use the following commands to implement the H.264 and H.265 coding schemes with the sequential-P mode and the hierarchical-B mode. 
Specifically, given a sequence \textit{Video.yuv} with the resolution as \textit{W×H}, the command lines for generating compressed video using x264 and x265 codecs are as follows: 
\begin{itemize}
\item H.264(P): \textit{ffmpeg -pix fmt yuv420p -s:v W×H -i Video.yuv -vframes $N_e$ -c:v libx264 -preset veryslow -x264-params “crf=CRF:keyint=GOP:bframes=0:scenecut=0” output.mkv}
\item H.264(B): \textit{ffmpeg -pix fmt yuv420p -s:v W×H -i Video.yuv -vframes $N_e$ -c:v libx264 -preset veryslow -x264-params “crf=CRF:keyint=GOP:scenecut=0:b-adapt=0:bframes=BF:b-pyramid=1” output.mkv}

\item H.265(P): \textit{ffmpeg -pix fmt yuv420p -s:v W×H -i Video.yuv -vframes $N_e$ -c:v libx265 -preset veryslow -x265-params “crf=CRF:keyint=GOP:bframes=0” output.mkv}
\item H.265(B): \textit{ffmpeg -pix fmt yuv420p -s:v W×H -i Video.yuv -vframes $N_e$ -c:v libx265 -preset veryslow -x265-params “crf=CRF:keyint=GOP:b-adapt=0:bframes=BF:b-pyramid=1” output.mkv}
\end{itemize}
where $N_e, CRF, BF$ represent the number of encoded frames, quantization parameter, and the number of B frames, respectively. Here we set $BF$ as GOP-1 for the hierarchical-B mode.

\subsection{Per-Video Level Analysis}
\label{appendix:rd_performance}
\begin{figure*}[t]
  \centering
  \subfloat 
  {\label{subfig:UVG_PSNR_details}
  \includegraphics[scale=0.315]{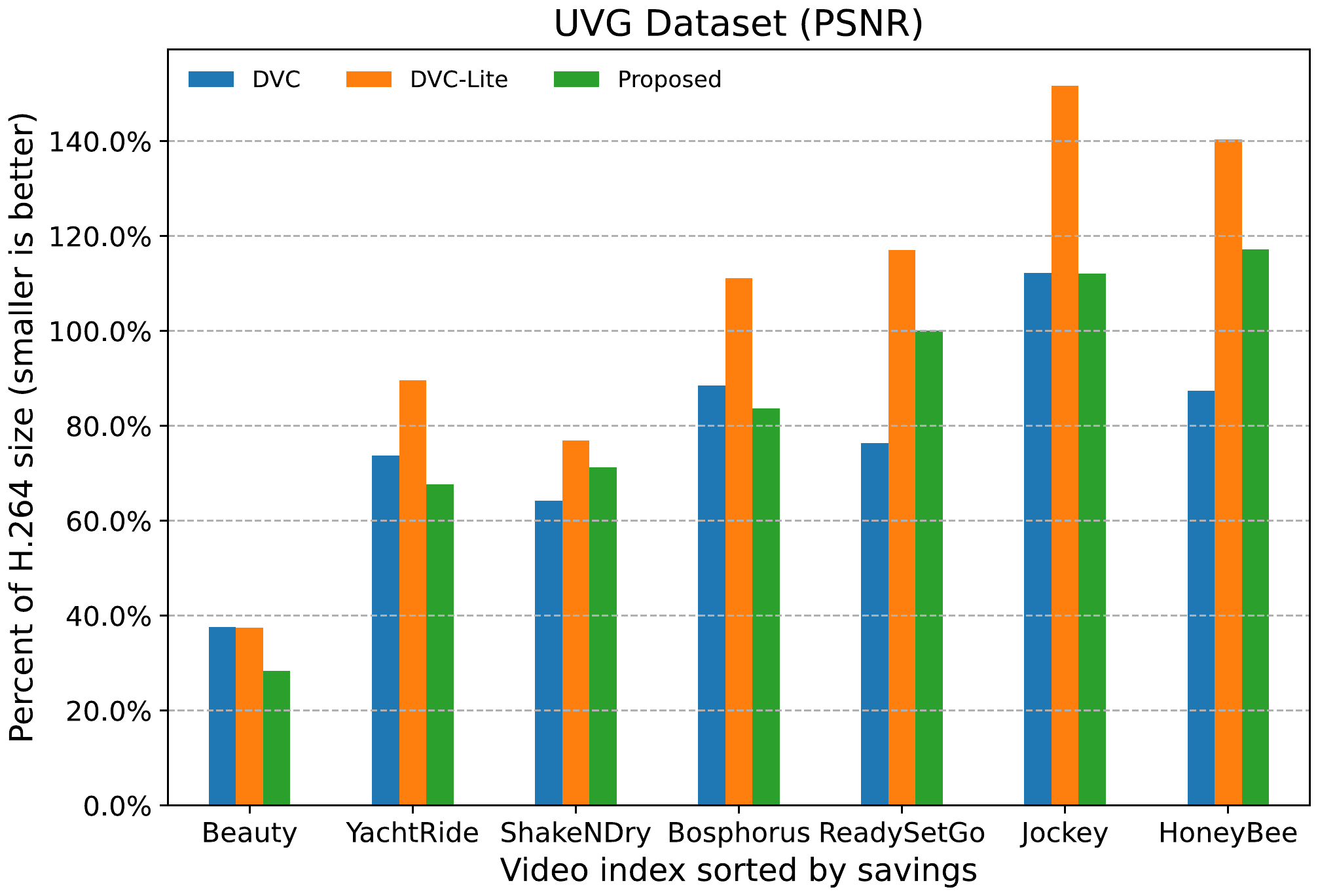}
  }
  \subfloat
  {\label{subfig:UVG_MS_SSIM_details}
  \includegraphics[scale=0.315]{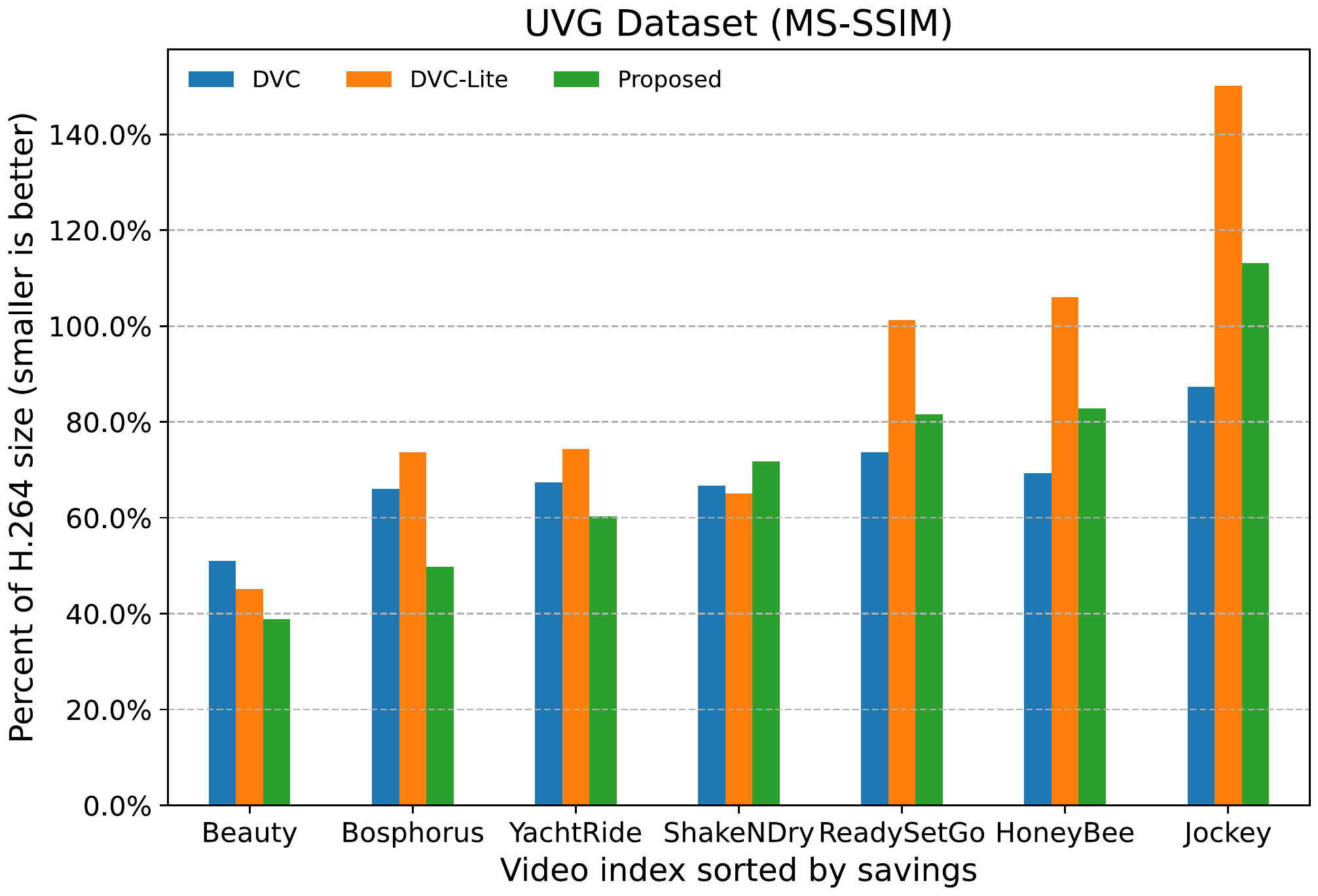}
  }\\
  \subfloat
  {\label{subfig:MCL_JCV_PSNR_details}
  \includegraphics[scale=0.35]{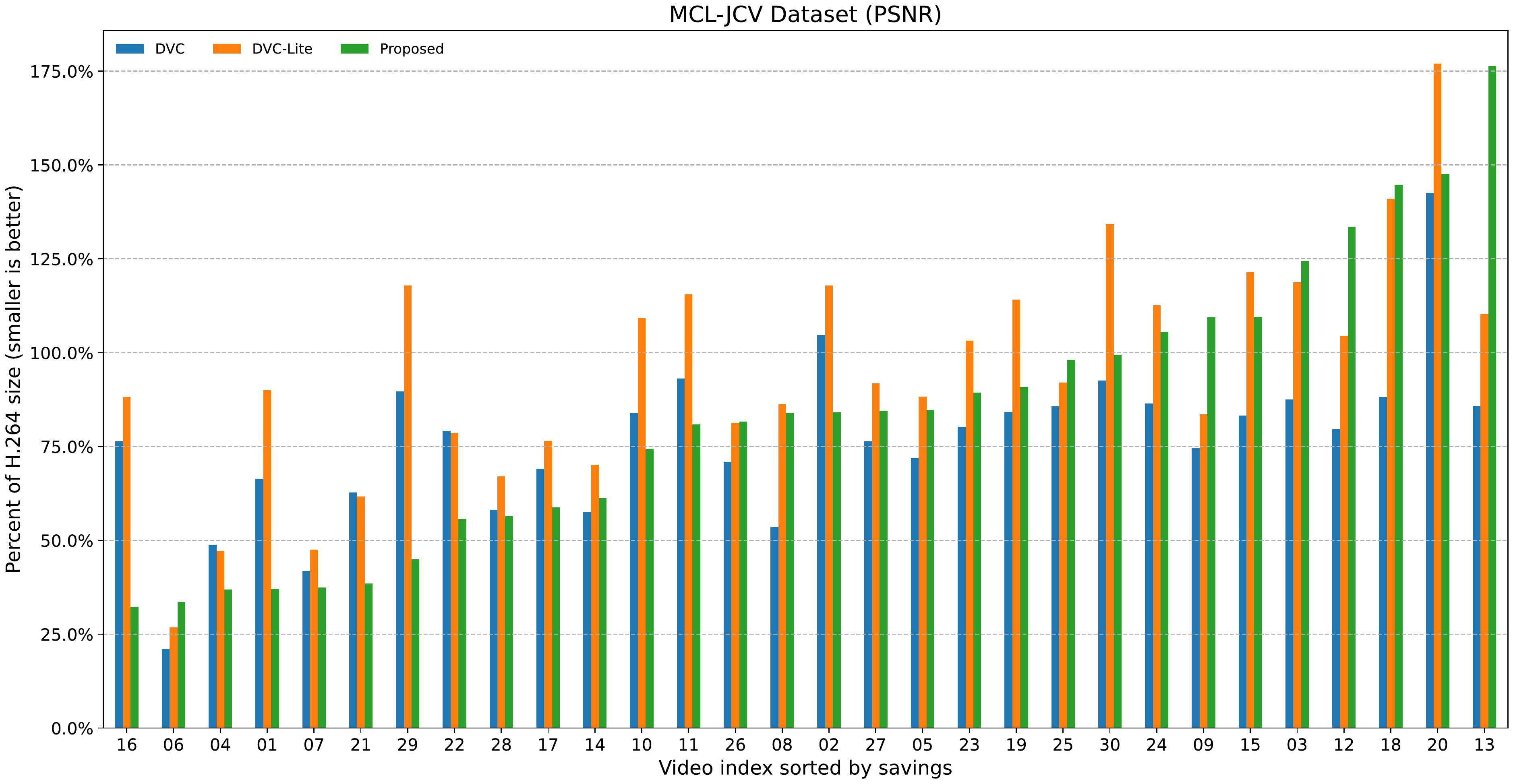}
  }\\
  \subfloat
  {\label{subfig:MCL_JCV_MS_SSIM_details}
  \includegraphics[scale=0.35]{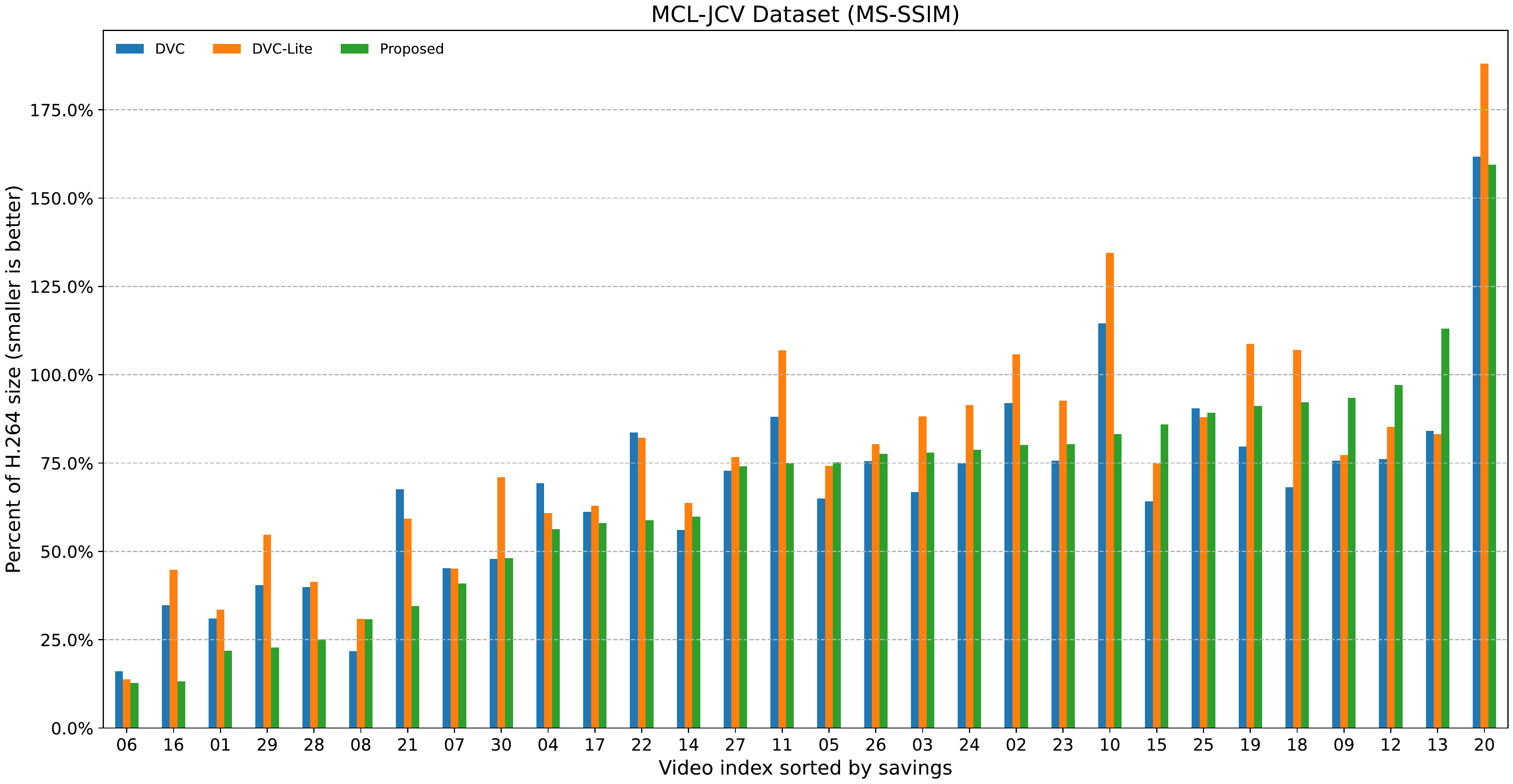}
  }
  \caption{Rate savings for each video on the UVG and MCL-JCV datasets. Values denote the relative size compared to H.264(P) when measured by BDBR at the same reconstruction level.}
  \label{fig:per_video_details}
\end{figure*}

In \figurename~\ref{fig:per_video_details}, H.264 (P) is set as the anchor to compute the BDBR for each video on the UVG and MCL-JCV datasets. We then plot the file size for each compressed video relative to H.264, \textit{e.g.}, a value of 30\% in the BDBR represents the relative size as 70\% (\textit{i.e.}, 100\%-30\%). 

It is observed that our proposed methods generate smaller encoded files than H.264 and DVC-Lite on most of videos. Compared to DVC, around half of the videos (16/37) compressed by the proposed method have smaller or equal file sizes in terms of PSNR. However, for a fraction of videos with fast motion information (\textit{e.g.}, ReadySetGo, video 03, 12 and 13), it is challenging for the proposed model to achieve the superior PSNR performance without utilizing temporal information at the encoder.
Moreover, due to the lack of animated videos in the training dataset, our methods cannot generalize well to cartoon videos (\textit{i.e.}, video 18, 20, 24, 25) in the MCL-JCV dataset.  

In addition, our methods have better compression efficiency in MS-SSIM than in PSNR, which is partly caused by exploiting SI in the feature space rather than pixel space at the decoder. Thus, the network inclines to focus on structure information instead of pixel information.
While there is some performance deficiency under the metric of PSNR in some videos, the proposed Distributed DVC framework is still a feasible attempt to approach the performance of predictive coding.



\subsection{Feature Visualizations}
\label{appendix:feature_visualizations}

\begin{figure*}[t]
  \centering
  \includegraphics[scale=0.38]{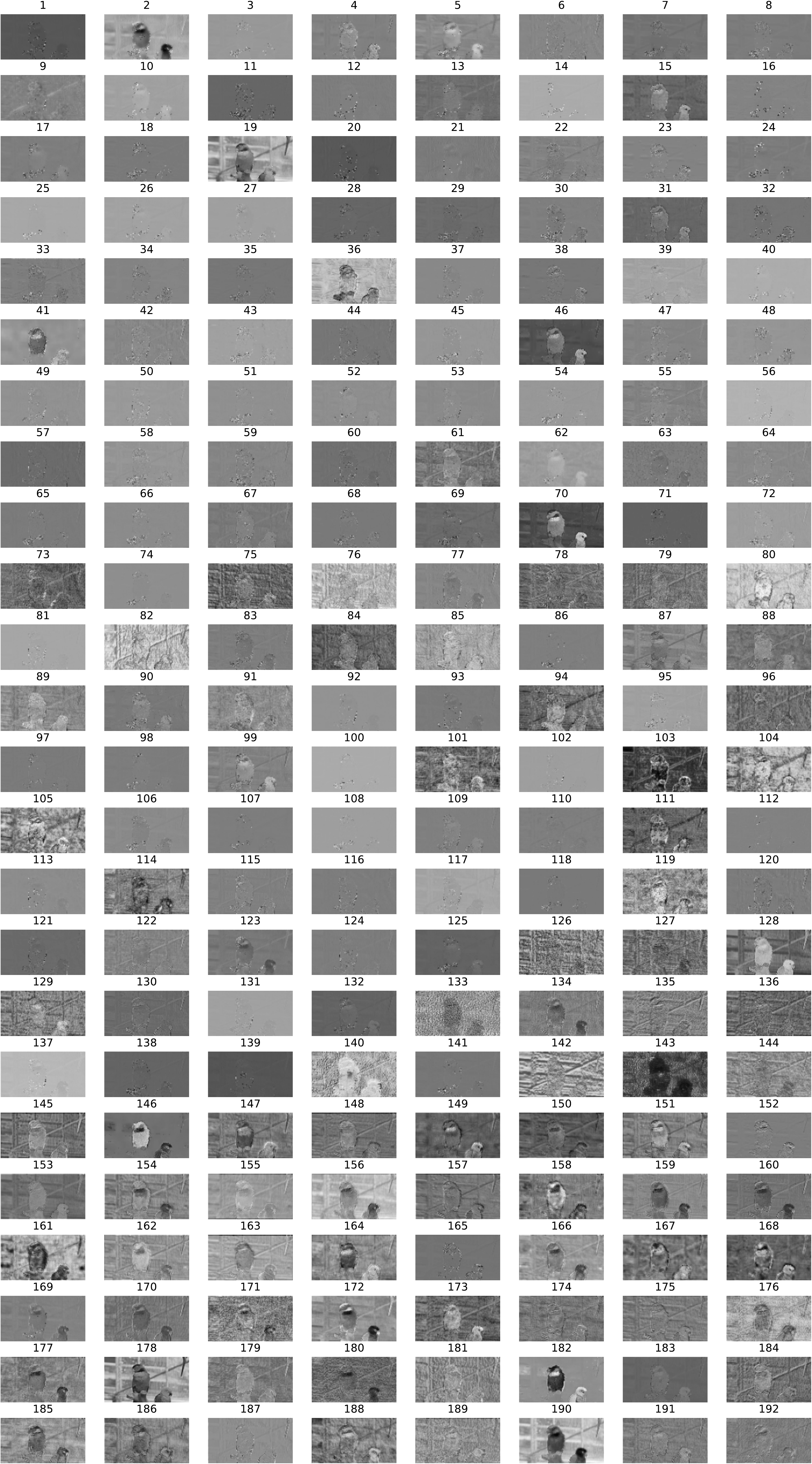}
  \caption{Visualizations of Wyner-Ziv representation from \textit{videoSRC30} sequence in the MCL-JCV dataset.}
  \label{fig:wz_representation}
  \end{figure*}

\begin{figure*}[t]
  \centering
  \includegraphics[scale=0.38]{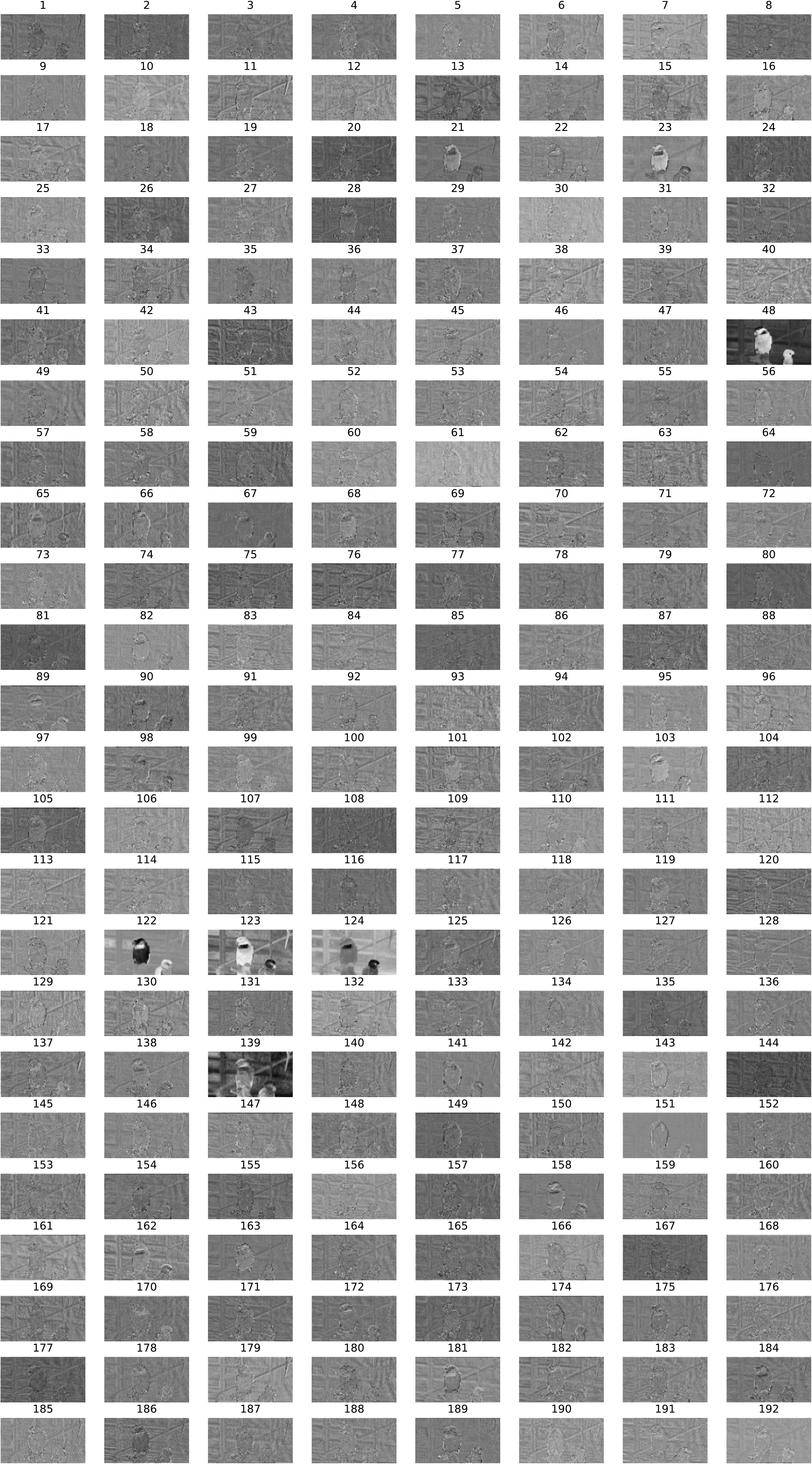}
  \caption{Visualizations of side information representation from \textit{videoSRC30} sequence in the MCL-JCV dataset.}
  \label{fig:si_representation}
  \end{figure*}

To better understand the difference between SI and WZ representations, we provide the visualization of WZ and SI feature maps in \figurename~\ref{fig:wz_representation} and \ref{fig:si_representation}. From the visualization results, we observe that when compared with the SI representation $\bar{y}_{kN+i}$, there are more (fewer) channels in $\hat{y}_{kN+i}$, e.g., the 128-th (30-th) and 178-th (170-th) channels, to represent low-frequency (high-frequency) information. Besides, only a few channels in $\bar{y}_{kN+i}$, e.g., the 48-th and 123-th channels, emphasize the low-frequency content, while most of channels of the SI representation, like the 151-th and 174-th channels, seem to pay more attention to the high-frequency details of the birds’ edges and silhouettes in contrast with high frequency in $x_{kN+i}$. This matches the reconstruction operation of the traditional WZ video coding where the high frequency DCT coefficients of the SI frame are directly used as that of the reconstructed WZ frame. These results suggest that it is worth exploring how to fully utilize the high-frequency information of SI feature and remove unimportant high-frequency feature maps in the WZ feature to improve the compression performance in the future. 


\end{document}